\documentstyle[12pt,epsfig]{article}

\newlength{\dinwidth}                                                    
\newlength{\dinmargin}
\def\lapproxeq{\lower .7ex\hbox{$\;\stackrel{\textstyle                                                    
<}{\sim}\;$}}                                                    
\def\gapproxeq{\lower .7ex\hbox{$\;\stackrel{\textstyle                                                    
>}{\sim}\;$}}                                                    
\def\be{\begin{equation}}                                                    
\def\ee{\end{equation}}                                                    
\def\bea{\begin{eqnarray}}                                                    
\def\eea{\end{eqnarray}}                                                    
\def\GeV{\rm GeV}
                                                    
\def\funp{{I\!\!P}}                    
                                                    
\begin{document}                                                    
\titlepage                                                    
\begin{flushright}                                                    
IPPP/09/19   \\
DCPT/09/38 \\                                                    
\today \\                                                    
\end{flushright}                                                    
                                                    
\vspace*{0.5cm}                                                    
                                                    
\begin{center}                                                    
{\Large \bf Forward Physics at the LHC\footnote{Based on a talk by Alan Martin at the Cracow Epiphany Conference on ``Hadron interactions at the dawn of the LHC, 5-7 January 2009; dedicated to the memory of Jan Kwiecinski''.}}                                                                                                        
                                                    
\vspace*{1cm}                                                    
A.D. Martin$^a$, M.G. Ryskin$^{a,b}$ and V.A. Khoze$^{a,b}$ \\                                                    
                                                   
\vspace*{0.5cm}                                                    
$^a$ Institute for Particle Physics Phenomenology, University of Durham, Durham, DH1 3LE \\                                                   
$^b$ Petersburg Nuclear Physics Institute, Gatchina, St.~Petersburg, 188300, Russia            
\end{center}                                                    
                                                    
\vspace*{1cm}                                                    
                                                    
\begin{abstract}
We review two inter-related topics. First, we consider the behaviour of ``soft'' scattering observables, such as $\sigma_{\rm tot}, ~d\sigma_{\rm el}/dt, ~d\sigma_{\rm SD}/dtdM^2,$ particle multiplicities etc., at high-energy proton-(anti)proton colliders.
We emphasize the sizeable effects of absorption on high-energy `soft' processes, and, hence, the necessity to include multi-Pomeron-Pomeron interactions in the usual multi-channel eikonal description. We describe a multi-component model which has been tuned to the available data for soft processes in the CERN-ISR to Tevatron energy range, and which therefore allows predictions to be made for `soft' observables at the LHC. The second topic concerns the calculation of the rate of exclusive processes of the form $pp \to p+A+p$ at high energy colliders, where $A$ is a heavy system. In particular, we discuss the survival probability of the rapidity gaps (denoted by the $+$ signs) to both eikonal and enhanced soft rescattering effects. At the LHC energy, the most topical case is when $A$ is a Higgs boson.  At the Tevatron, measurements have been made for the exclusive diffractive production of various systems: $A$ being either $\gamma\gamma$, dijet, or a $\chi_c$  
meson. We compare the observed rates with the expectations. Finally, we describe how predictions for exclusive processes may be checked in the early runs of the LHC. 

\end{abstract}    

\section{Motivation for revisiting soft interactions}

The description of the high energy behaviour of ``soft'' scattering observables such as $\sigma_{\rm tot}$, $d\sigma_{\rm el}/dt$,   $d\sigma_{\rm SD}/dtdM^2,$ particle multiplicities etc. at hadron colliders predated QCD. It is sometimes regarded as the Dark Age of strong interactions. However, this is not totally fair. There was a successful description of these processes in terms of the exchange of Regge trajectories linked to particle states in the crossed channels \cite{reviews}. The dominant exchange at high energy is the Pomeron, and we have Gribov's Reggeon calculus \cite{RFT} to account for the multi-Pomeron contributions. However the available data did not reach high enough energy to distinguish between the different scenarios \cite{GM1,GM2} for the high-energy
behaviour of the interaction amplitude \cite{MGR,KMR-08}.

With the advent of the LHC, it is valuable to revisit soft interactions\footnote{Two papers resulted from Jan Kwiecinski's first extended visit to Durham in 1990. These were entitled ``Partons at small $x$''\cite{jk1} and ``Semihard QCD expectations for $pp$ (or $p\bar{p}$) scattering at CERN, the Tevatron and the SSC''\cite{jk2}. It is interesting to observe that the first rapidly developed into a rich and very enjoyable collaborative programme, with impact on HERA, whereas the second was almost 20 years premature.} because of
\begin{itemize}
\item the intrinsic interest in obtaining a reliable, self-consistent model for soft interactions, which may be illuminated by data from the LHC; 
\item the need to use the model for predictions of the gross features of soft interactions; in particular, to understand the structure of the underlying events at the LHC;
\item the advantages of using central exclusive production, $pp \to p+H+p$, to study the properties of the Higgs sector in an especially clean environment at the LHC
\cite{KMR}~-~\cite{CR1}.
 For example, the Higgs mass(es) can be measured with very good accuracy
($\Delta M_H\sim 1-2$ GeV) by the missing-mass method by detecting the
outgoing very forward protons.  
However, the exclusive cross sections are strongly suppressed by the small
survival factor, $S^2 \ll 1$, of the rapidity gaps. Thus we need a reliable model of soft interactions to evaluate
the corresponding value of $S^2$ \cite{oldsoft,RMK2}.

\end{itemize}

\section{Description of soft interactions}

\begin{figure}
\begin{center}
\includegraphics[height=6cm]{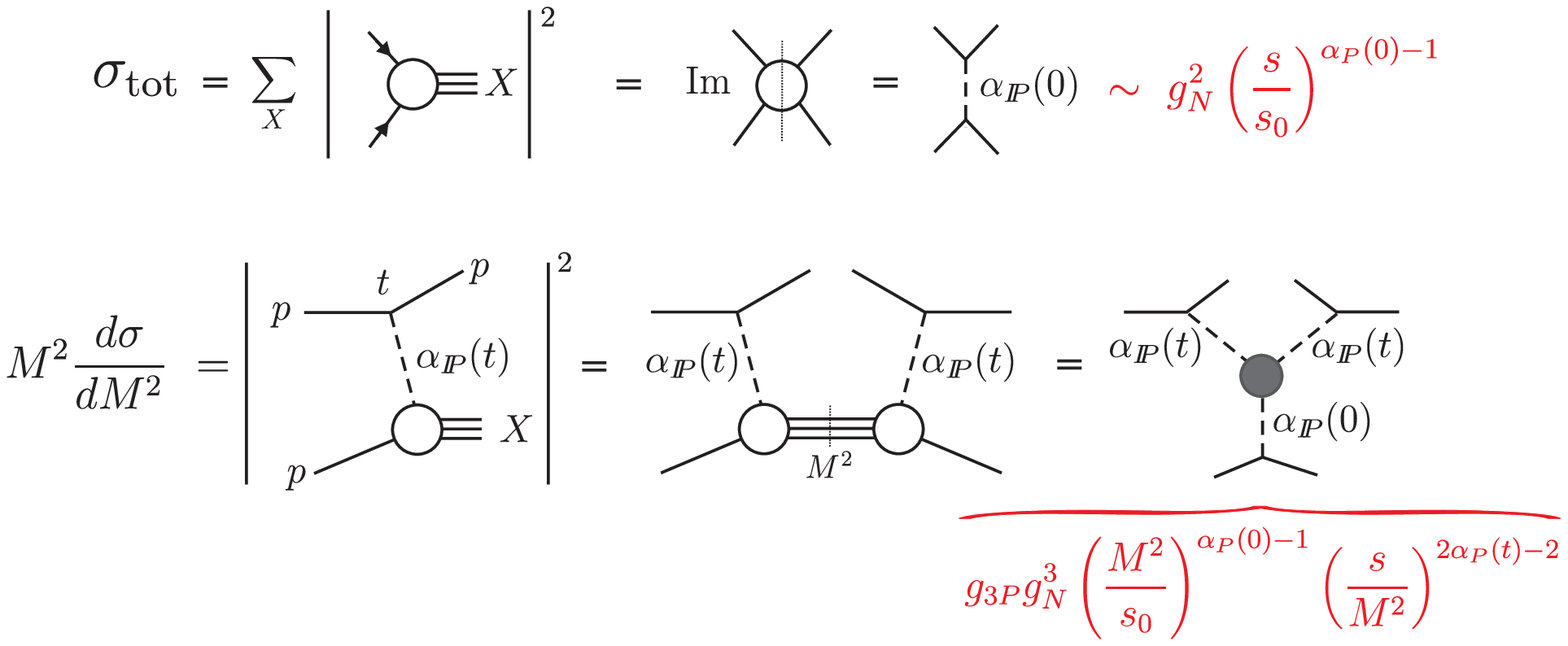}
\caption[*]{Optical theorems for the total cross section and for high-mass diffractive dissociation. The high-energy Regge expressions shown for the {\it bare} amplitudes have sizeable absorptive corrections.}
\label{fig:opt}
\end{center}
\end{figure}

We start with Fig.~\ref{fig:opt}, which shows sketches of the optical theorems for the total cross section and for high-mass single diffractive dissociation. The high-energy expressions given for the cross sections arise from the Regge behaviour of the {\it bare} amplitudes.  However there are screening (that is, absorptive) corrections, which will suppress the cross sections. For example, the value of the triple-Pomeron coupling, $g_{3P}$, originally obtained \cite {FF} using the {\it bare} formula,
 will, on taking account of the absorptive corrections \cite{LKMR} , be increased.

\subsection{Single channel eikonal $-$ elastic scattering}

\begin{figure} [h]
\begin{center}
\includegraphics[height=8cm]{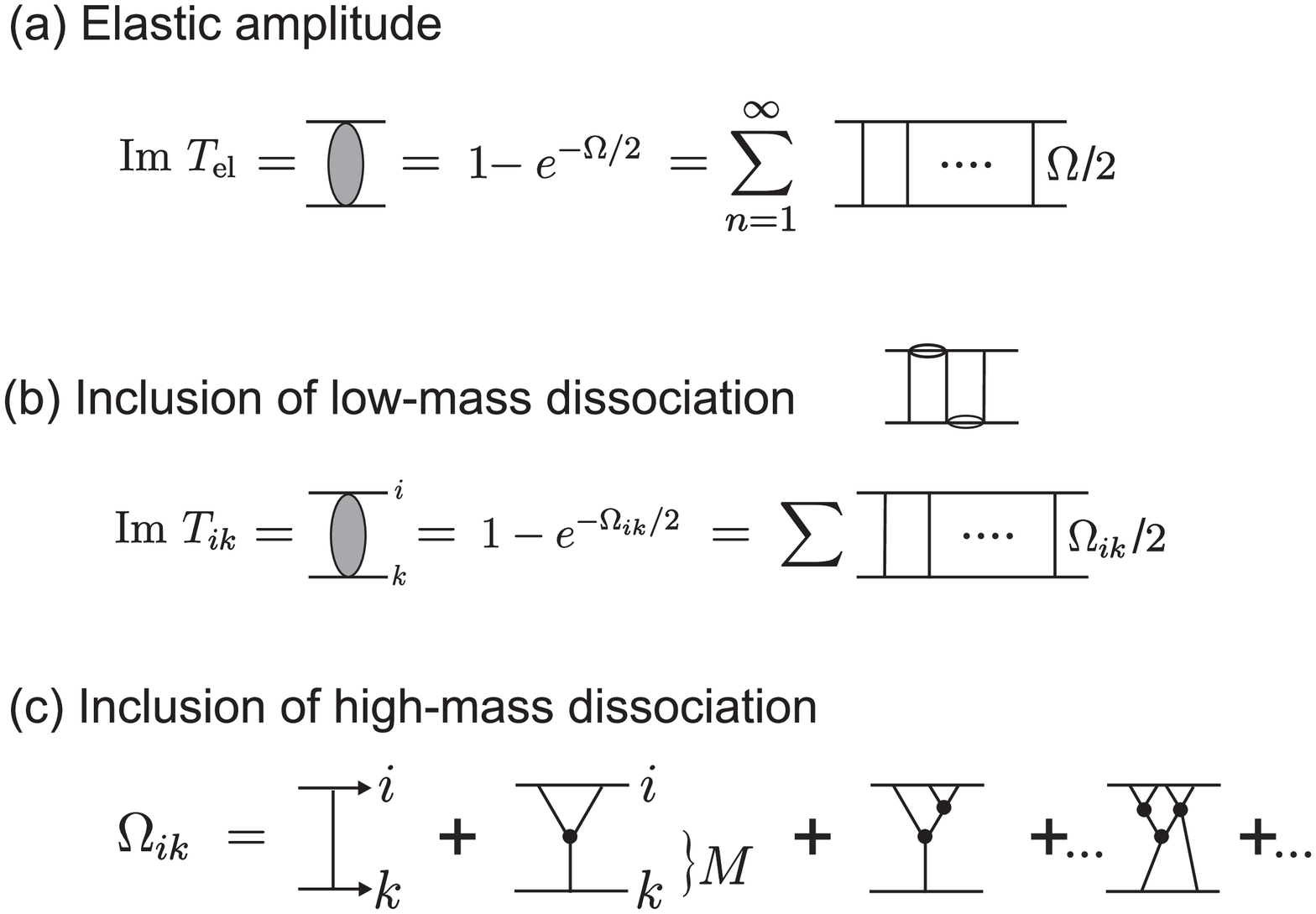}
\caption[*]{(a) The single-channel eikonal description of elastic scattering; (b) the multichannel eikonal formula which allows for low-mass proton dissociations in terms of diffractive eigenstates $|\phi_i\rangle,~|\phi_k\rangle$; and (c) the inclusion of the multi-Pomeron-Pomeron diagrams which allow for high-mass dissociation. }
\label{fig:AA}
\end{center}
\end{figure}

The total and elastic cross sections are usually described in terms of an eikonal model. At high energy the position of the fast particle in the impact parameter, $b$, plane is, to a good approximation, frozen during the interaction. Thus the value of $b$ is related to the orbital angular momentum $l=b\sqrt{s}/2$ of the incoming hadron. Elastic unitarity, therefore, takes the form\footnote{$G_{\rm inel}$ accounts for the presence of inelastic channels.} 
\be 2 {\rm Im}\,T_{\rm el}(s,b) = |T_{\rm el}(s,b)|^2 + G_{\rm
inel}(s,b),
 \label{eq:a1} 
\ee
from which it follows that
\bea
\frac{d \sigma_{\rm tot}}{d^2 b} & = & 2  {\rm Im}T_{\rm el}(s,b)  = 2( 1-{\rm e}^{-\Omega/2}) \label{eq:elastamp}\\
\frac{d \sigma_{\rm el}}{d^2 b} & = & |T_{\rm el}(s,b)|^2  =  (1-{\rm e}^{-\Omega/2})^2, \label{eq:el}\\
\frac{d \sigma_{\rm inel}}{d^2 b} & = & 2  {\rm Im}T_{\rm el}(s,b) - |T_{\rm el}(s,b)|^2 =  1- {\rm e}^{-\Omega}, \label{eq:inel} 
\eea
where $\Omega(s,b)\geq 0$ is called the
opacity (optical density) or eikonal\footnote{Sometimes $\Omega/2$
is called the eikonal.}.  It is the Fourier transform of the two-particle ($s$-channel)
{\it irreducible} amplitude, $A(s,q_t)$. That is\footnote{We use the bold face symbols ${\mathbf q}_t$ and ${\mathbf b}$ to denote vectors in the transverse plane.}
\begin{equation}
\label{eq:append9}
\Omega (s, b) ~=~\frac{-i}{4\pi^2}\int d^2
q_t~A(s,q_t)e^{i{\mathbf q}_t\cdot{\mathbf b}} \; ,
\end{equation}
 where $q_t^2=-t$, and where the amplitude is normalized by the relation
$\sigma_{\rm tot}(s)=2{\rm Im}T_{\rm el}(s,t=0)$.
>From (\ref{eq:inel}), we see that $ \exp(-\Omega(s,b))$ is the probability that
no inelastic scattering occurs at impact parameter $b$. 
In the framework of the eikonal model, the elastic amplitude,
is obtained by the sum of Regge-exchange diagrams, as shown in Fig.~\ref{fig:AA}(a), which is equivalent to the iteration of the elastic unitarity equation, (\ref{eq:a1}).

  At high energies the ratio ${\rm Re} T_{\rm el}/{\rm Im} T_{\rm el}$ is small and can be
evaluated via the dispersion relation. So the imaginary part of $\Omega$
is usually neglected.

\subsection{Inclusion of diffractive dissociation}
So much for elastic diffraction. Now we turn to inelastic
diffraction, which is a consequence of the {\em internal
structure} of hadrons. Besides the pure elastic two-particle intermediate states shown in Fig.~\ref{fig:AA}(a), there is the possibility of proton excitation, $p \to N^*$. At high
energies, where the lifetime of the fluctuations of the fast proton is
large, $\tau\sim E/m^2$, the
corresponding Fock states can be considered as `frozen'. Each constituent of the proton can undergo scattering and thus destroy the
coherence of the fluctuations. As a consequence, the outgoing
superposition of states will be different from the incident
particle, so we
will have {\em inelastic}, as well as elastic, diffraction.

To discuss inelastic diffraction, it is convenient to follow Good
and Walker~\cite{GW}, and to introduce states $|\phi_k \rangle$ which
diagonalize the $T$ matrix. Such, so-called diffractive, eigenstates only undergo elastic
scattering.  To account for the internal structure of the hadronic states, we, therefore, have to enlarge the set of
intermediate states, from just the single elastic channel, and to
introduce a multichannel eikonal. The situation is pictured in Fig.~\ref{fig:AA}(b). It is straightforward to show that 
\be
\label{eq:b6} \frac{d \sigma_{\rm tot}}{d^2 b} ~ = ~ 2 \langle T \rangle ~~~~~~~~{\rm and}~~~~~~~~\frac{d \sigma_{\rm el}}{d^2 b}~ = ~ \langle T \rangle^2 
\ee
 where the
brackets of $\langle T \rangle$ mean that we take the average of
${\rm Im}T(s,b)$ over the initial probability distributions of diffractive
eigenstates of the `beam' and `target' protons.
The cross section for the single diffractive
dissociation of the `beam' proton,
\be \label{eq:b7} \frac{d \sigma_{\rm SD}}{d^2 b} \; = \; \langle
(T)^2 \rangle \: - \: \langle T \rangle^2, \ee
is given by the statistical dispersion in the absorption
probabilities of the diffractive eigenstates. Here, the average is only
taken over the diffractive components into which the `beam' proton
dissociates.

At first sight, it appears that if we were to enlarge the number of eigenstates $|\phi_i\rangle$, then we may include even high-mass proton dissociation. However here we face the problem of double counting when partons originating from dissociation of the beam and `target' initial protons overlap in rapidities. For this reason high-mass dissociation is usually described by ``enhanced'' multi-Pomeron diagrams. The first and simplest is the triple-Pomeron graph, shown in Fig.~\ref{fig:opt}, and again in Fig.~\ref{fig:AA}(c).

\subsection{The importance of absorptive effects}
 Already at Tevatron energies the absorptive
correction to the elastic amplitude, due to elastic eikonal
rescattering, is not negligible; it is about $-20$\% in comparison with the
simple one Pomeron exchange. After accounting for low-mass proton
excitations (that is $N^*$'s in the intermediate states) the correction becomes twice
larger (that is, about $-40$\%). Indeed, the possibility of proton excitation means that we have to include additional inelastic channels which were not accounted for in the irreducible amplitude $A$ of (\ref{eq:append9}).  This enlarges the probability of absorption for the elastic channel, that is the effective opacity $\Omega$. 

Next, when we account for the screening of high-mass diffractive dissociation, $d\sigma_{\rm SD}/dM^2$, there is an extra factor
of 2 coming from the AGK cutting rules \cite{AGK}. Thus, the absorptive effects
in the triple-Regge domain are expected to be quite large. The previous
triple-Regge analyses (see, for example, \cite{FF}) did not allow for
absorptive corrections and the resulting triple-Regge couplings must be
regarded, not as {\it bare} vertices, but as {\it effective} couplings
embodying the absorptive effects \cite{capella}. Since  the inelastic cross section
(and, therefore, the absorptive corrections) expected at the LHC are more
than twice as large as that observed at fixed-target and CERN-ISR
energies,  the old triple-Regge vertices cannot be used to predict the
diffractive cross sections at the LHC.

Thus, it was necessary  to perform a new triple-Regge analysis
that includes the absorptive effects explicitly.
Such an analysis has recently been performed \cite{LKMR} in which the fixed-target FNAL, CERN-ISR and Tevatron data, that are
available in the triple-Regge region, are fitted in terms of `$PPP$', `$PPR$', `$RRP$', `$RRR$' and $\pi\pi P$ contributions\footnote{The analysis of \cite{LKMR} had the limited objective of estimating the triple-Reggeon couplings; in particular the triple-Pomeron coupling $g_{3P}\equiv g_{PPP}$. It should not be used as a model to describe soft high energy interactions in the whole rapidity interval. Section \ref{sec:model} describes such a model.}.  To account for the absorptive
corrections a two-channel (Good-Walker) eikonal model was used, which describes well the
total, $\sigma_{\rm tot}$, and elastic, $d\sigma_{\rm el}/dt$, $pp$ and $\bar
pp$ cross sections.

Since the absorptive effects are included explicitly,
   the extracted values of the triple-Reggeon vertices are now much closer to the {\it
bare} triple-Regge couplings. The value obtained,
\begin{equation}
g_{3P}\, \equiv \, \lambda g_N,~~~~~~~~~{\rm with}~~\lambda \simeq 0.2,
\label{eq:2}
\end{equation}
is about a factor of 3 larger than that found in the original analyses, which did not include absorptive corrections.  Here, $g_N$ is the Pomeron-proton coupling. The new value of the coupling, $g_{3P}$, is consistent with a reasonable extrapolation of the perturbative BFKL
Pomeron vertex to the low scale region \cite{BRV}.  Note also that the resulting
values of the `$PPP$' and `$PPR$' vertices allow a good description
\cite{KMRpl}  of the HERA
data \cite{Jpsi} on inelastic $J/\psi$ photoproduction, $\gamma p\to
J/\psi+Y$, where the screening corrections are rather small.

\subsection{Multi-component $s$- and $t$-channel model of soft processes \label{sec:model}}

Since the effects due to the triple-Pomeron vertex (\ref{eq:2}) are rather large, we must include all the multi-Pomeron diagrams, some of which are shown in Fig.~\ref{fig:AA}(c). Why do we claim a large effect when $\lambda$ is only 0.2? The reason is that the contribution caused by such vertices is enhanced by the logarithmically large phase space available in rapidity. In particular, the total cross section of high-mass dissociation is roughly\footnote{Here, for simplicity, we assume an essentially flat energy dependence, $\sigma \sim s^{\epsilon}$ with $\epsilon {\rm ln}s <1$. The final equality in (\ref{eq:SDel}) can be deduced by taking the ratio of the couplings indicated in the Regge expressions in Fig.~\ref{fig:opt}.} of the form
\be
\sigma_{\rm SD}~=~\int \frac{M^2 d\sigma_{\rm SD}}{dM^2}~\frac{dM^2}{M^2}~\sim~\lambda{\rm ln}s~\sigma_{\rm el},
\label{eq:SDel}
\ee
where $\lambda$ reflects the suppression of high-mass dissociation in comparison with elastic scattering and the ln$s$ factor comes from the integration $\int dM^2/M^2~\sim~{\rm ln}s$. Thus actually we deal with the parameter $\lambda {\rm ln}s \gapproxeq 1$ at collider energies. For each fixed rapidity interval the probability of high-mass dissociation (or, in other words, the contribution due to the triple-Pomeron vertex) is relatively small. However the cumulative effect in the complete interaction amplitude is {\rm enhanced} by the large phase space available in rapidity.

As a consequence, the contribution of the corresponding, so-called `enhanced', diagrams, with a few
vertices,
  is not negligible. Moreover, we expect that more complicated
multi-Pomeron interactions, driven by the $g^n_m$ vertices (which
describe the transition of $n$ to $m$ Pomerons), will affect the
final result.  Certainly\cite{gnm}, it is more reasonable to assume that $g^n_m\propto
\lambda^{n+m}$ than to assume that $g^n_m=0$ for any $n+m>3$. Thus, we
need a model which accounts for the possibility of multi-Pomeron interactions
(with arbitrary $n$ and $m$).

\begin{figure}
\begin{center}
\includegraphics[height=3.15cm]{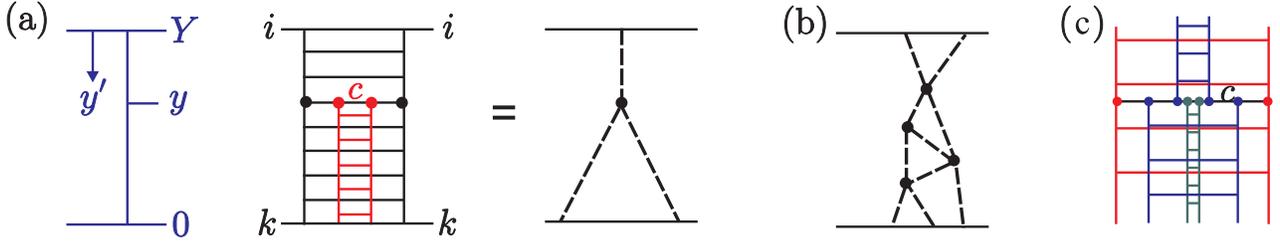}
\caption{(a) The ladder structure of the triple-Pomeron amplitude between diffractive eigenstates $|\phi_i\rangle,|\phi_k\rangle$ of the proton; the rapidity $y$ spans an interval 0 to $Y={\rm ln}s$. (b) A multi-Pomeron diagram.  (c) The ``nested'' ladder structure of the $g^2_3$ multi-Pomeron vertex.} 
\label{fig:2lad}
\end{center}
\end{figure}
Here we follow the {\it partonic} approach of Refs.~\cite{KMRs1,KMRnns}.
While the eikonal formalism describes the rescattering of the incoming
fast particles,  the enhanced multi-Pomeron diagrams
represent the rescattering of the intermediate partons in the ladder
(Feynman diagram) which describes the Pomeron-exchange amplitude.
We refer to Fig.~\ref{fig:2lad}. The multi-Pomeron effects are included by the following equation describing the evolution in rapidity $y$ of the opacity $\Omega_k$ starting from the `target' diffractive eigenstate $|\phi_k\rangle$:
\begin{equation}
\frac{d\Omega_k(y,b)}{dy}\,=\,e^{-\lambda\Omega_i(y',b)/2}~~~e^{-\lambda
\Omega_k(y,b)/2}~~
\left(\Delta+\alpha'\frac{d^2}{d^2b}\right)\Omega_k(y,b)\; ,
\label{eq:evol1}
\end{equation}
where $y'=\ln s -y$.   Let us explain the meanings of the three factors on the right-hand-side of (\ref{eq:evol1}). If only the last factor, (...)$\Omega_k$, is present then the evolution generates the ladder-type structure of the bare Pomeron exchange amplitude, where the Pomeron trajectory $\alpha_P=1+\Delta+\alpha't$. The inclusion of the preceding factor allows for rescatterings of an intermediate parton $c$ with the ``target'' proton $k$; Fig.~\ref{fig:2lad}(a) shows the simplest (single) rescattering which generates the triple-Pomeron diagram. Finally, the first factor allows for rescatterings with the beam $i$. In this way the absorptive effects generated by all multi-Pomeron diagrams are included, like the one shown in Fig.~\ref{fig:2lad}(b). 
There is an analogous equation for the evolution in rapidity $y'$ of $\Omega_i(y',b)$ starting from the `beam' diffractive eigenstate $|\phi_i\rangle$. The two equations may be solved iteratively.

As we are dealing with elastic {\it amplitudes} we use $e^{-\lambda\Omega/2}$ and not $e^{-\lambda\Omega}$. The coefficient $\lambda$ in the exponents arises since  parton $c$ will have a different absorption cross section from that of the diffractive eigenstates. Naively, we may assume that the states $i,k$ contains a number $1/\lambda$ of partons.  The factors $e^{-\lambda\Omega/2}$  generate multi-Pomeron vertices of the form 
\begin{equation}
g^n_m\,=\, n~m~\lambda^{n+m-2}g_N/2\;\;\;\;\;\;\;\;\;
\mbox{for~~ $n+m\geq 3$}\, ,
\label{eq:g}
\end{equation}
where a factor $1/n!$, which comes from the expansion of the exponent, accounts for the
identity of the Pomerons. The factors $n(m)$ allow for the $n(m)$ possibilities to select the Pomeron $\Omega_i(\Omega_k)$ which enters the evolution (\ref{eq:evol1}) from the $n(m)$ identical Pomerons.  

Even though $\lambda \simeq 0.2-0.25$, the role of factors $e^{-\lambda\Omega/2}$ is not negligible, since the suppression effect is accumulated throughout the evolution. For instance, if $\lambda \ll 1$ the full absorptive correction is given by the product $\lambda \Omega Y/2$, where the small value of $\lambda$ is compensated by the large rapidity interval $Y$.

Strictly speaking nothing is known, either experimentally or
theoretically, about the behaviour of multi-Pomeron
vertices, $g_m^n$, at low scales. At large scales, for a large
number of colours, $N_c$, the leading contribution is given by
diagrams like Fig.~\ref{fig:2lad}(a,c), which contain the $g^1_2$ and $g^2_3$ multi-Pomeron vertices respectively, where the Pomeron ladders are `nested' within each other.
Here the factor $1/n!$ arises from the time ordering (that is, from the
ordering of the longitudinal momentum fractions) in the cells of
each Pomeron adjacent to the parton $c$ line; the lifetime of an {\it inner} Pomeron should be
smaller than that for an {\it outer} Pomeron.
Thus, in the large $N_c$ limit, the form (\ref{eq:g}) of the
multi-Pomeron vertices, with $\lambda\propto N_c\alpha_s/\pi$, is
motivated by perturbative QCD. At moderate values of $N_c$, unfortunately there
are other, more complicated, contributions. In particular, the $2 \to 2$ Pomeron
vertex $g^2_2$ has a term proportional to $\alpha_s/(N_c^2-1)^2$,
that is to the first power of $\alpha_s$ \cite{lrs}, and not the third power of $\alpha_s$ as in (\ref{eq:g}). (Note that, in perturbative QCD, $g_N\propto \alpha_s$ as well.)
However, this term is strongly suppressed by the colour factor
$1/(N^2_c-1)^2=1/64$ and, most probably, such terms are not too
important at low scales, when the coupling $\alpha_s$ is not very
small.

So far, we have allowed multi-components in the $s$-channel via a multichannel eikonal. However,
a novel feature of the model of Ref.~\cite{KMRnns} is that four different $t$-channel states are included. One for the
secondary Reggeon ($R$) trajectory, and three Pomeron states ($P_1, P_2, P_3$) to mimic the BFKL
diffusion in the logarithm of parton transverse momentum,
$\ln(k_t)$ \cite{Lip}. Recall that the
BFKL Pomeron \cite{bfkl} is not a pole in the complex $j$-plane, but a
branch cut. Here the cut is approximated by three $t$-channel states of a different size.
The typical values of $k_t$ are $k_{t1}\sim 0.5$ GeV, $k_{t2}\sim 1.5$ GeV and $k_{t3}\sim 5$ GeV for the large-, intermediate- and small-size components of the Pomeron, respectively.
Thus (\ref{eq:evol1}) is rewritten as a four-dimensional matrix equation for $\Omega^a_k$ in $t$-channel space ($a=P_1,P_2,P_3,R$), as well as being a three-channel eikonal in diffractive eigenstate $|\phi_k\rangle$ space. The transition terms, added to the equations, which couple the different $t$-channel components, are fixed by the properties of the BFKL equation. So, in principle, we have the possibility to explore the matching of the soft Pomeron (approximated by the large-size component $P_1$) to the QCD Pomeron (approximated by the small-size component $P_3$). The key parameters which drive the evolution in rapidity are the intercepts $1+\Delta^a$ and the slopes $\alpha'_a$ of the $t$-channel exchanges.

Section \ref{sec:excl} will be devoted to the exclusive central production of different heavy systems $A$. That is, to processes of the type $pp \to p+A+p$.
In Section \ref{sec:survival}, we will see that the multi-component Pomeron will allow us to obtain an estimate of the survival probability, $S^2_{\rm enh}$, of the rapidity gaps (denoted by the $+$ signs) to so-called {\it enhanced} soft rescattering; that is, to rescattering on {\it intermediate} partons with different $k_t$. Clearly enhanced rescattering will violate soft-hard factorisation of the process. Note that $S_{\rm enh}$ depends on $k_t$, which is driven by the scale of the central hard subprocess. We need to introduce components of the Pomeron of different size to be able to calculate $S_{\rm enh}$ for different central exclusive processes, such as $A$ = Higgs, dijet, $\gamma\gamma$ or $\chi_c$.
These enhanced rescattering effects are in addition to the survival of the gaps to {\it eikonal} soft rescattering, which preserves soft-hard factorisation.

\subsection{Description of the `soft' data and predictions for the LHC}

The number of parameters in the model of Ref.~\cite{KMRnns} is too large to perform a straightforward $\chi^2$ fit of the data. Instead, the
majority of the parameters are fixed at physically reasonable values, and it was
demonstrated that all the features of the available data on
diffractive cross sections, $\sigma_{\rm tot},\, d\sigma_{\rm el}/dt,\,
\sigma_{\rm SD}^{{\rm low}M},\,
d\sigma_{\rm SD}/dM^2$, in the CERN-ISR to Tevatron range are reproduced.

\begin{figure} [t] 
\begin{center}
\includegraphics[height=12cm]{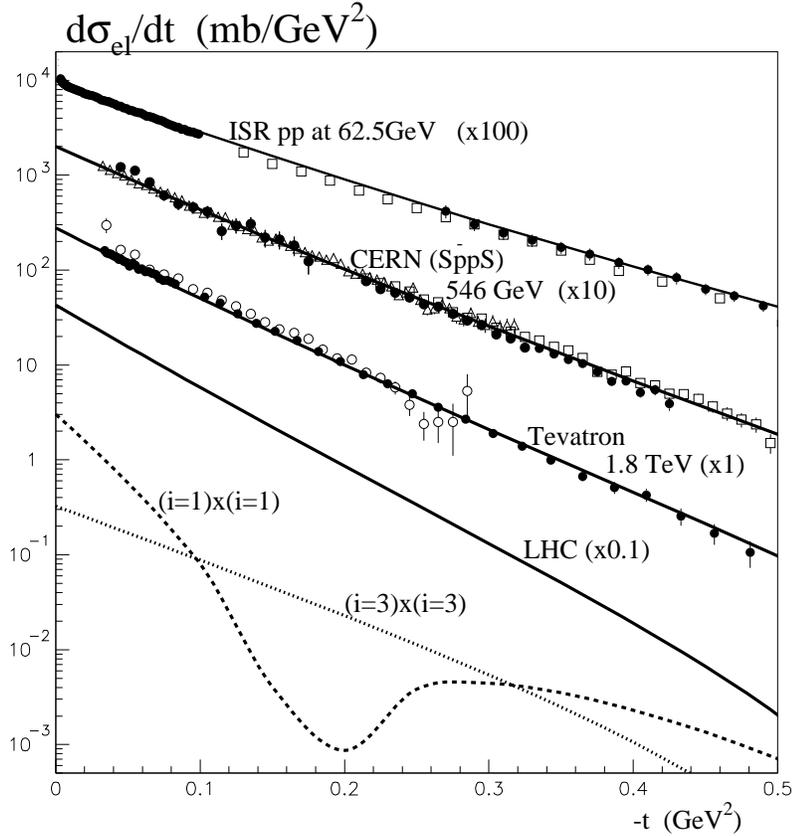}
\caption[*]{The $t$ dependence of the elastic $pp$ cross section \cite{KMRnns}. The
dashed and dotted lines are the contributions from  the elastic
scattering of the largest size ($i=1$) and the smallest size ($i=3$)
diffractive components.} 
\label{fig:dsdt}
\end{center}
\end{figure}

Fig.~\ref{fig:dsdt} shows the quality of the description \cite{KMRnns} of the 
data for the {\it elastic differential} cross section. 
We also present in Fig.~\ref{fig:dsdt} the prediction for differential elastic cross section at
the LHC energy $\sqrt{s}=14$ TeV.
Recall that \cite{KMRnns} uses a three-channel eikonal. That is $i,k=1,2,3$. It is interesting to note that the contribution to the cross section arising from the scattering of the two large-size diffractive
eigenstates, $(i=1)\times(i=1)$, already has a diffractive dip at $-t=0.2$ GeV$^2$. However, after the contributions from all
possible combinations $i\times k$ are summed up, the
prediction has no dip up to $-t=0.5$ GeV$^2$. 

\begin{figure} [h,t]
\begin{center}
\includegraphics[height=12cm]{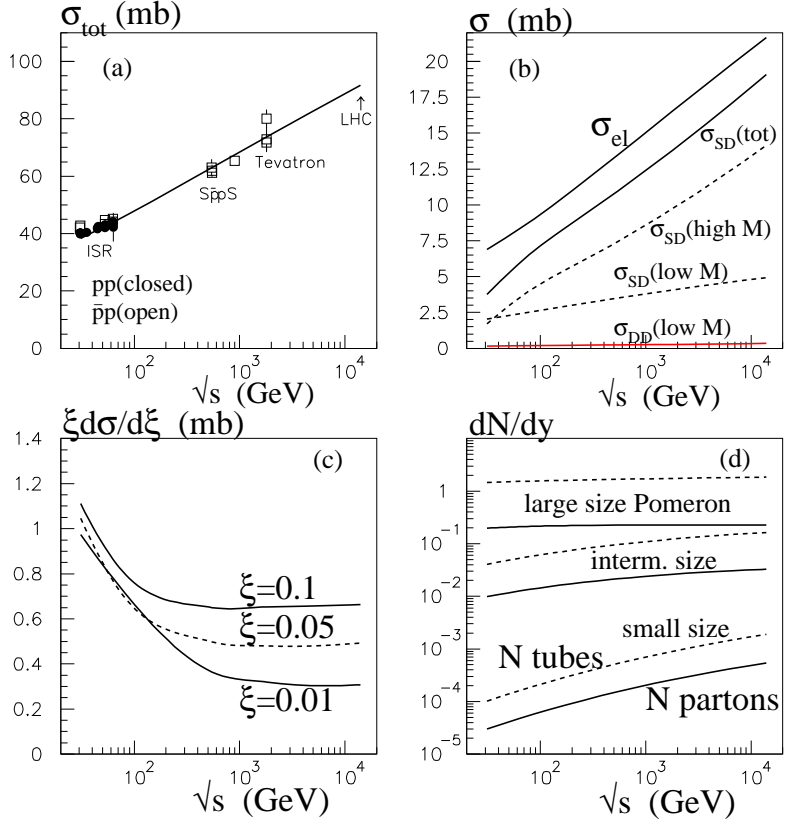}
\caption{The energy dependence of the total (a), elastic and
diffractive dissociation (b) $pp$ cross sections and the cross sections of
dissociation to a fixed $M^2=\xi s$ state (c); (d) the parton
multiplicity (solid lines) and the number of `colour  tubes' (dashed)
produced by the Pomeron components of different size. The figure is taken from \cite{KMRnns}.}
\label{fig:sum4}
\end{center}
\end{figure}

To describe the high-energy behavior of the {\it total} cross section, $\Delta^a\equiv \alpha^a(0)-1=0.3$ is taken for each of the three components of the Pomeron.
These Pomeron intercepts are consistent with resummed NLL BFKL,
which gives $\omega_0\sim 0.3$ practically independent of the scale
$k_t$ \cite{BFKLnnl}. The slopes of the Pomeron trajectories are driven by the transverse momentum associated with the particular component $a$; in fact, $\alpha'\propto 1/k^2_t$.
The data require $\alpha'_{P_1}=0.05$ GeV$^{-2}$ for the large-size Pomeron component, so in the model we take
$\alpha'_{P_2}=0.05/9$ GeV$^{-2}$ for the second component and $\alpha'_{P_3}=0$ for the smallest-size component.
For the secondary Reggeon trajectory the model takes $\alpha'_R=0.9$ GeV$^{-2}$,
and $\alpha_R(0)=0.6$. The `bare' intercept is a bit larger than $\frac{1}{2}$,
since the final effective intercept is reduced by the absorptive
corrections included in the evolution equation. The description of the total cross section data are shown in Fig.~\ref{fig:sum4}(a). The screening corrections arising from the `enhanced' multi-Pomeron
  diagrams, that is from the high-mass dissociation, slow down the growth of
  the cross section with energy.  Thus, the model predicts a  relatively low
total
cross section at the LHC -- $\,\sigma_{\rm tot}({\rm LHC})\simeq 90$
mb\footnote{Low values, $\sim$90 mb, are also predicted by other models of `soft' interactions which include absorptive effects \cite{KGB,GLMM}.}. 

\begin{figure} [h]
\begin{center}
\includegraphics[height=7cm]{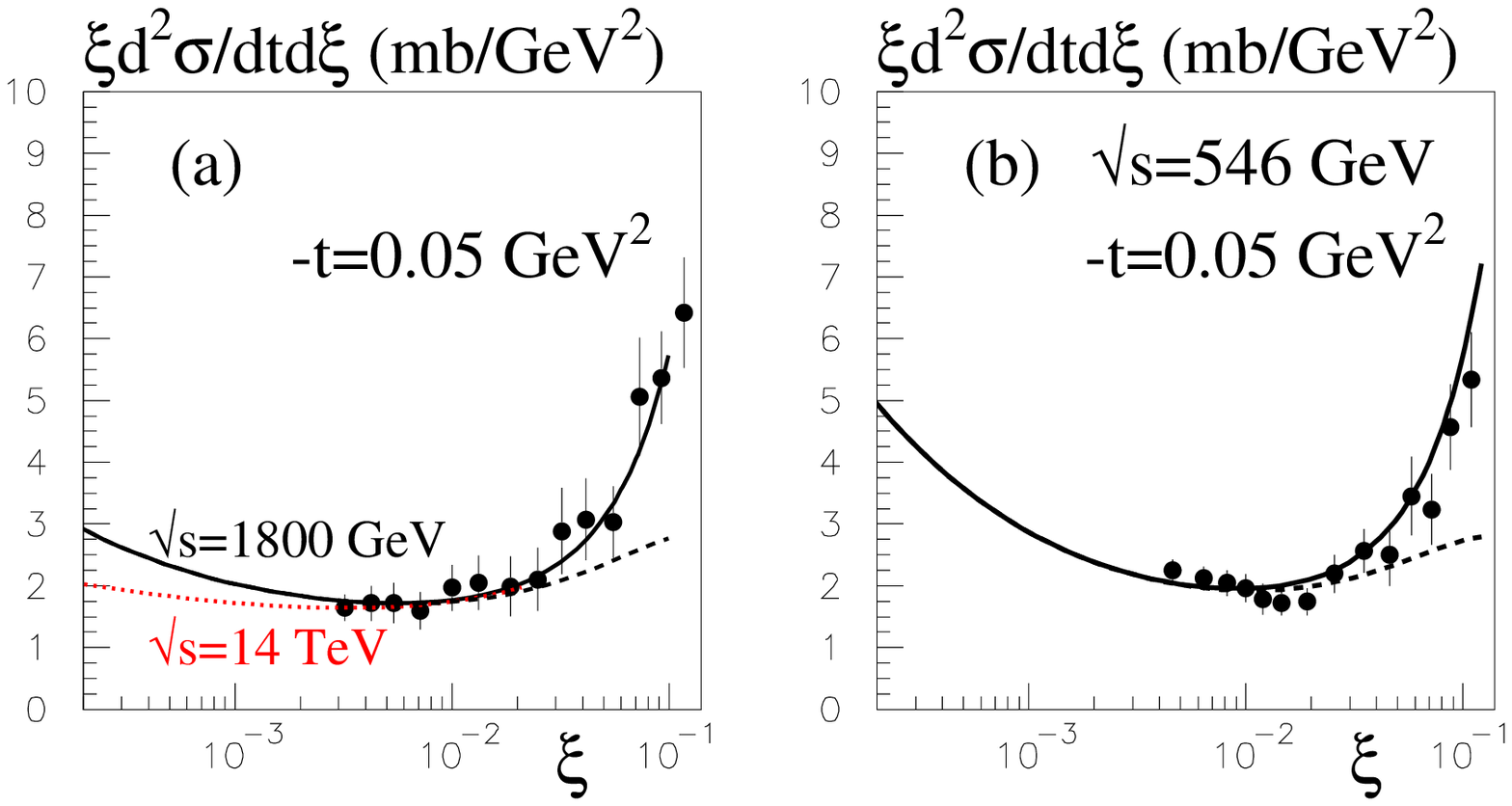}
\caption{The model description \cite{KMRnns} of the data for the cross section for high-mass dissociation versus $\xi$ for $-t=0.05~{\rm GeV}^2$ at $\sqrt{s}=$ 1800 GeV and 546 GeV \cite{CDFhm,GoulMon}. The dashed lines are the predictions without the $\pi\pi P$ contribution. The dotted curve at small $\xi$ is the prediction for the LHC.}
\label{fig:dsy}
\end{center}
\end{figure}

The value of the parameter $\lambda$, which controls the cross section of {\it high-mass
dissociation} in the small $\xi$ (that is, large $y$) region, was found to be $\lambda \simeq 0.25$.
The dependence of the cross section for high-mass dissociation, $\xi d^2\sigma/dtd\xi$, on $\xi=M^2/s$ is compared with the Tevatron CDF data \cite{CDFhm,GoulMon} in Fig.~\ref{fig:dsy}.  The results without the $\pi\pi P$ term are shown by the dashed lines. We also show in Fig.~\ref{fig:dsy}(a) (by the dotted line at small $\xi$) the prediction for the LHC energy. 

The energy behaviour of the cross sections are shown in Table 1 and Fig.~\ref{fig:sum4}. We do not quote the values of the cross section for double diffractive dissociation, $\sigma_{\rm DD}$. The model of \cite{KMRnns} will give results similar to those in Table 2 of Ref.~\cite{KMRs1}. These values of $\sigma_{\rm DD}$ are in excellent agreement with the cross sections observed at the Tevatron.  
\begin{table}[htb]
\begin{center}
\begin{tabular}{|l|c|c|c|c|c|}\hline
energy &   $\sigma_{\rm tot}$ &  $\sigma_{\rm el}$ &    $\sigma_{\rm SD}^{{\rm low}M}$ &  $\sigma_{\rm SD}^{{\rm high}M}$  &   $\sigma_{\rm SD}^{\rm tot}$ \\ \hline

 1.8  &   73.7   &     16.4  &       4.1      &   9.7  &     13.8  \\
 14    &  91.7   &     21.5     &    4.9   &     14.1  &     19.0    \\
 100   &  108.0  &     26.2  &       5.6   &     18.6  &     24.2   \\ \hline

\end{tabular}
\end{center}
\caption{Cross sections (in mb) versus collider energy (in TeV) \cite{KMRnns}.}
\end{table}

The values of $\sigma_{\rm SD}^{\rm tot}$ quoted in Table 1 look, at first sight, too large, when compared with the value  $9.46 \pm 0.44$ mb given by CDF \cite{CDFhm}. However the CDF value does not include the secondary Reggeon ($RRP$) contribution, denoted as a `non-diffractive' component of $2.6 \pm 0.4$ mb. Moreover, the trigger used to select the diffractive dissociation events rejects part of the low-mass proton excitations. Taking these absences into account, there is no contradiction between the model prediction and the CDF data. Furthermore, note that in the region where the CDF detector efficiency and resolution are good, our model gives a good description of the differential cross sections which were actually measured, see  Fig.~\ref{fig:dsy}.

As we have a detailed model for high energy soft processes, it would appear to be possible to predict the multiplicity distribution at the LHC. However, although some general features can be predicted, unfortunately, we cannot reliably calculate the
multiplicity distributions of secondary {\it hadrons}, since
 a non-negligible fraction of the final hadrons may be produced via the
fragmentation of minijets.   We  can only discuss
 the multiplicity distribution at the partonic level.

\begin{figure} [h] 
\begin{center}
\includegraphics[height=12cm]{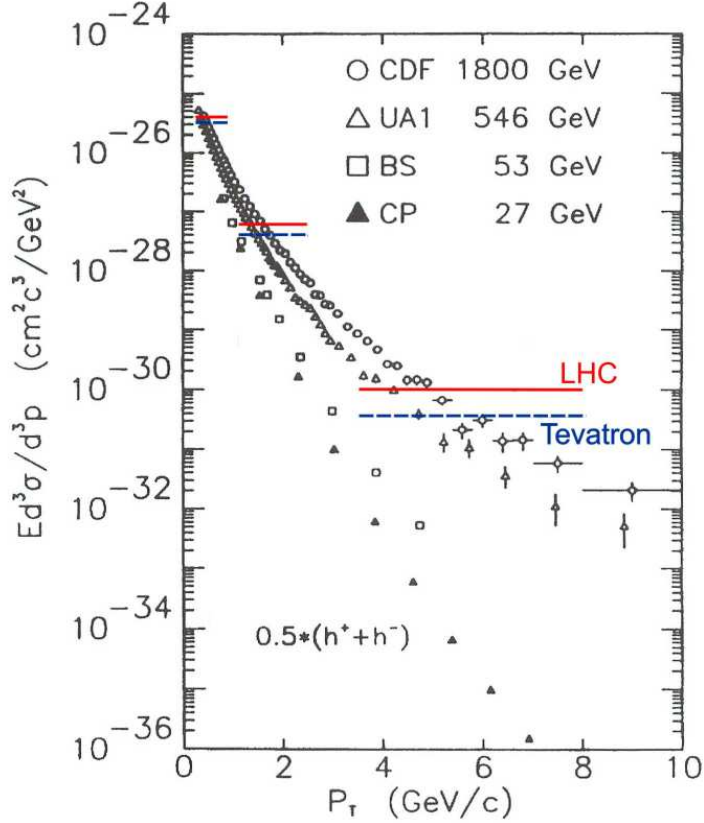}
\caption[*]{The plot is from Ref.~\cite{CDF61}. The horizontal lines, which are superimposed, are the predictions of \cite{KMRnns} at the Tevatron and LHC energies; the three $p_t$ ranges correspond to the large-, intermediate- and small-size components of the Pomeron.} 
\label{fig:inclmult}
\end{center}
\end{figure}

 The mean number of the ($t$-channel) ladders of the type $a$
 produced in the collision of $i$ and $k$ Good-Walker eigenstates
 can be calculated. 
 This quantity may be considered as the mean number of colour tubes
 of type $a$ produced in the proton-proton interaction.
 To obtain the number of partons created by the ladder `$a$' at 
 rapidity $y$, we have to allow for  
 the parton density $\rho^a(y)$. 
The results are shown in Fig.~\ref{fig:sum4}(d).  The main growth in multiplicity, as we go from Tevatron to LHC energies, is due to the small size (`QCD') Pomeron component, which produces particles with typically $p_t \sim 5$ GeV. There is essentially no growth in multiplicity at small $p_t$. This simply confirms the trend that has been observed through the CERN-ISR to Tevatron energy range, see the data points in Fig.~\ref{fig:inclmult}.

In other words, starting with the same intercepts ($\Delta=0.3$) the contribution of the large-size
component after the absorptive correction becomes practically
flat, while  the small-size contribution, which is much less affected by
the absorption, continues to grow with energy. As mentioned above, such a behaviour is
consistent with the  experiment (see Fig.~\ref{fig:inclmult}) where the density of low $k_t$
secondaries is practically saturated while probability to produce a hadron
with a large (say, more than 5 GeV) transverse momentum grows with the
initial energy.
 To obtain a qualitative feel for the expected behaviour at the hadronic level, we show the estimates of \cite{KMRnns} at the Tevatron and LHC energies in Fig.~\ref{fig:inclmult}, where the horizontal lines indicate the typical $p_t$ interval associated with each Pomeron component.

\section{Exclusive processes at hadron colliders \label{sec:excl}}

In this Section we discuss exclusive processes of the type $pp \to p+A+p$ where the $+$ signs denote large rapidity gaps. The mechanism for the process in sketched in Fig.~\ref{fig:pAp}. 
Although, below, we will consider several different heavy systems $A$, the main motivation is the possibility to use the process $pp \to p+H+p$ to detect and study one or more Higgs bosons at the LHC.

\subsection{Advantages of the exclusive Higgs signal}

\begin{figure}
\begin{center}
\includegraphics[height=4cm]{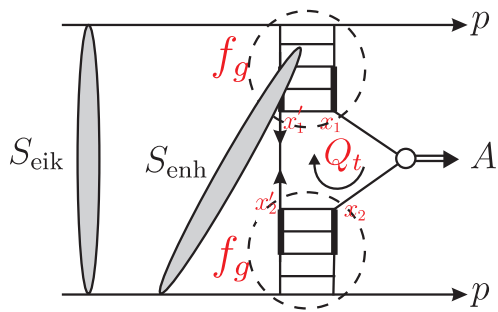}
\caption{The mechanism for the exclusive process $pp \to p+A+p$, with the eikonal and enhanced survival factors shown symbolically. The thick lines on the Pomeron ladders, either side of the subprocess ($gg \to A$), indicate the rapidity interval $\Delta y$ where enhanced absorption is not permitted, see Section \ref{sec:survival}. }
\label{fig:pAp}
\end{center}
\end{figure}

We consider the mass range $M_H \lapproxeq 140$ GeV, where the dominant decay mode is $H \to b\bar{b}$. 
The exclusive process has a small cross section;
$\sigma_{\rm excl}~\sim ~10^{-4}~\sigma^{\rm tot}_{\rm incl}$ at the LHC energy.
Nevertheless, it has the following advantages:
\begin{itemize}
\item
The mass of the Higgs boson (and in some cases the width) can be measured
with high accuracy (with mass resolution $\sigma(M)\sim 1$ GeV) by measuring the
missing mass to the forward outgoing protons, {\it provided} that they can be accurately tagged some 400 m from the interaction point
\cite{FP420}.
\item
The leading order $b\bar b$
  QCD background is suppressed \cite{Jz} by the P-even $J_z=0$ selection
rule\footnote{$J_z=0$ originates from $s$-channel helicity conservation of the
forward protons if their transverse momenta 
$p_{ti}=0$, with $i=1,2$.   The admixture of the $|J_z|=2$ {\it amplitudes} is
governed by the product $({\mathbf p}_{t1} \cdot {\mathbf r}) ~({\mathbf p}_{t2} \cdot {\mathbf r}) ~\simeq ~ \langle p_t^2 \rangle/Q_t^2$,
where the transverse size of the digluon Pomeron $r \sim 1/Q_t$, and $Q_t$ is the transverse
momentum in the gluon loop in Fig.~\ref{fig:pAp}. Thus, the $|J_z|=2$ contribution to the cross section
is suppressed by a factor $p_t^4/Q_t^4$.  The presence of the Sudakov form factor
provides the infrared stability of the $Q_t$ integral over the gluon loop.
Typically, the main contribution comes from the saddle-point region of
the loop integral, namely $Q_t^2 \sim 4~\GeV^2$
for the exclusive production of a scalar Higgs of mass $M_H=120$ GeV at the LHC energy,
and so ensures that $J_z=0$ is a very good approximation.
The P even selection reflects the symmetry between the active gluons emitted from
the two incoming protons.}, where the $z$ axis is along the direction of the proton beam.
Indeed, at LO, this background vanishes in the limit of massless $b$ quarks and forward outgoing protons.
Therefore, one can observe the Higgs boson via the main decay mode $H\to
b\bar b$, which otherwise is very challenging to measure in inclusive production due to the overwhelming QCD background. Moreover, a measurement of the mass of the decay products must match the `missing mass' measurement. 
\item
The quantum numbers of the central object (in particular, the
C- and P-parities) can be analysed by studying the azimuthal angle
distribution of the tagged protons. Due to the selection
rules, the production of $0^{++}$ states is strongly favoured.
\item
There is a clean environment for the
exclusive process -- the soft background is strongly suppressed \cite{KMRinsight}.
\item
Extending the study to BSM Higgs bosons
\cite{KKMRext}~-~\cite{katri}.
For example,
there are regions of MSSM parameter space were the
signal is enhanced by a factor of 10 or more, while the background remains unaltered
\cite{KKMRext,hkrstw,clp}. Moreover,
there are domains of parameter space where MSSM Higgs boson production via the conventional inclusive processes
 is suppressed whereas the exclusive signal is
enhanced, and even such, that both the $h$ and $H$ $0^{++}$ bosons may be detected \cite{hkrstw}.   
\end{itemize}

The main background to the exclusive $H \to b\bar{b}$ signal comes from (i) irreducible QCD $gg^{PP} \to b\bar{b}$ events, (ii) gluons mimicking $b$ jets and (iii) the $|J_z|=2$ contribution arising from non-forward going protons. The superscript $PP$ is to indicate that the active gluons originate within overall colourless hard Pomeron exchanges. Recently
\cite{shuv}, the NLO contribution to (i) has been calculated and reduces the irreducible background by a factor of 2 or more.
 Radiation from the screening gluon in Fig.~\ref{fig:pAp} is numerically small \cite{myths}.
 In summary, it should be possible to achieve a $H \to b\bar{b}$ signal-to-background ratio of $O(1)$, and much greater in some BSM Higgs scenarios
\cite{hkrstw,hkrtw,katri}.

The downside is that the exclusive cross section is small: about 3 fb for a 120 GeV SM Higgs at the LHC. Allowing for acceptance cuts, $b$ jet recognition etc., it corresponds to only a handful, or so, observable events for an integrated luminosity of 30 fb$^{-1}$. Moreover, there are still some experimental issues to be settled; in particular, very good timing is necessary to identify an exclusive event from among  the multiple interactions per bunch crossing which will occur at high LHC luminosities.

\subsection{Calculation of the exclusive Higgs signal}

The calculation of the exclusive production of a heavy system is an interesting mixture of {\it soft} and {\it hard} QCD effects. Here we concentrate on exclusive scalar Higgs production, $pp \to p+H+p$. 
The basic mechanism is
shown in Fig.~\ref{fig:pAp}. The $t$-integrated cross section is of the form
\begin{equation}
\sigma ~\simeq ~\frac{S^2}{B^2} ~\left|~N\int\frac{dQ^2_t}{Q^4_t}\: f_g(x_1, x_1', Q_t^2, \mu^2)f_g(x_2,x_2',Q_t^2,\mu^2)~\right|^2, 
\label{eq:M}
\end{equation}
where $B/2$ is the $t$-slope of the proton-Pomeron vertex, and $N$ is given in terms of the $H\to gg$ decay width.
The probability amplitudes, $f_g$, to find the appropriate pairs of $t$-channel gluons $(x_1,x'_1)$ and $(x_2,x'_2)$, are given by the skewed unintegrated gluon densities at a hard scale $\mu \sim M_H/2$. Since $(x'\sim Q_t/\sqrt s)\ll (x\sim M_H/\sqrt s)\ll 1$, it is possible
to express $f_g(x,x',Q_t^2,\mu^2)$, to single log accuracy\footnote{This can be achieved, to a good approximation, by setting the lower limit of the $k_t$ integration in the Sudakov form factor equal to $Q_t$ \cite{KMRearly}, based on the BFKL equation; and the upper limit
 $\mu =0.62M_H$. The factor 0.62 follows from an exact calculation of the one-loop contribution \cite{KKMRext}. }, in
terms of the conventional integrated density $g(x)$, together with a known Sudakov suppression factor $T$, which ensures that the active gluons do not radiate in the
evolution from $Q_t$ up to the hard scale $\mu \sim M_H/2$, and
so preserves the rapidity gaps. The factor $T$ ensures that the integral is infrared stable, and may be calculated by perturbative QCD.

The factor $S^2$ in (\ref{eq:M}) is the probability that the secondaries, which are produced by soft rescattering do not populate the rapidity gaps. It is the price we have to pay for an exclusive process. As written, the cross section assumes soft-hard factorization. It assumes that the survival factor, denoted by $S_{\rm eik}$ in Fig.~\ref{fig:pAp} and calculated from a model of soft interactions, does not depend on the structure of the perturbative QCD amplitude embraced by the modulus signs in (\ref{eq:M}). Actually the situation is more complicated, see, for instance, \cite{RMK2,KMRearly}.

\subsection{Rapidity gap survival \label{sec:survival}}

The gap survival factor caused by {\it eikonal} rescattering of the
diffractive eigenstates \cite{GW}, for a fixed impact parameter ${\mathbf b}$, is
\begin{equation}  S^2_{\rm eik}({\mathbf b})~ = ~\frac{\left| {\displaystyle\sum_{i,k}} |a_{i}|^2~|a_{k}|^2~{\mathcal
M}_{ik}({\mathbf b})~\exp(-\Omega^{\rm tot}_{ik}(s,{\mathbf b})/2)\right|^2}{\left|{\displaystyle \sum_{i,k}}
|a_{i}|^2~|a_{k}|^2~{\mathcal M}_{ik}({\mathbf b}) \right |^2} \,.
\label{eq:c3pp}
\end{equation}
where $\Omega^{\rm tot}_{ik}(s,{\mathbf b})$ is the total opacity of the $ik$ interaction, and the $a_i$'s occur in the decomposition of the proton wave function  in terms of diffractive eigenstates $|p\rangle = \sum_i a_i |\phi_i \rangle$.  The total opacity has the form $\Omega^a_k(y)\Omega^a_i(y')$ integrated over the impact parameters ${\mathbf b_1},{\mathbf b_2}$ (keeping a fixed impact parameter separation ${\mathbf b}={\mathbf b_1}-{\mathbf b_2}$ between the incoming protons) and summed over the different Pomeron components $a$. Recall $y'=Y-y={\rm ln}s-y$, see Fig.~\ref{fig:2lad}.  The exact shape of the matrix element ${\cal M}_{ik}$ for the hard subprocess $gg \to H$ in ${\mathbf b}$ space and the relative couplings to the various diffractive eigenstates $i,k$ should be addressed further.

\begin{figure}
\begin{center}
\includegraphics[height=7cm]{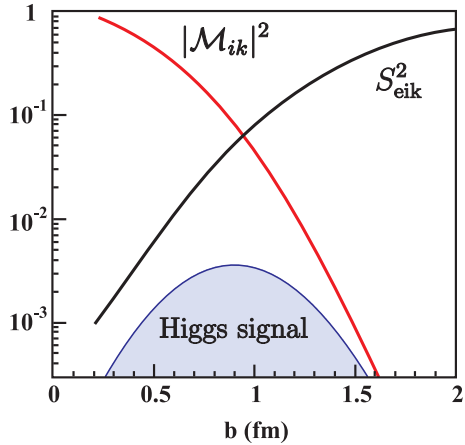}
\caption{A ``first look'' at the impact parameter dependence of the signal for 120 GeV Higgs production at the LHC after including an eikonal rescattering correction. }
\label{fig:higgs}
\end{center}
\end{figure}

One possibility is to say that the ${\mathbf b}$ dependence of ${\cal M}$
should be, more or less, the same as for diffractive
$J/\psi$ electroproduction ($\gamma+p\to J/\psi+p$), and the coupling
to the $|\phi_i \rangle$ component of the proton should be proportional to the same factor 
$\gamma_i$ as in a soft interaction. This leads to
\begin{equation}
{\cal M}_{ik}\propto \gamma_i\gamma_k\exp(-b^2/4B)
\label{eq:m1}
\end{equation}
with $t$-slope $B\simeq 4$ GeV$^{-2}$ \cite{JPsi}.
The resulting ``first look'' predictions obtained using the `soft' model of \cite{KMRnns}, for the exclusive production of a scalar 120 GeV Higgs at the LHC, are shown in Fig.~\ref{fig:higgs}. After we integrate over $b$, we find that the survival probability of the rapidity gaps in $pp \to p+H+p$ to eikonal rescattering is $\langle S^2_{\rm eik}\rangle$=0.017, with the Higgs signal concentrated around impact parameter  $b=0.8$ fm.
Expressing the survival factors in this manner is too simplistic and even sometimes misleading, for the reasons we shall explain below; nevertheless these numbers are frequently used as a reference point. 

As indicated in Fig.~\ref{fig:pAp}, besides {\it eikonal} screening, $S_{\rm eik}$, caused by soft interactions between the protons, we must also consider so-called {\it `enhanced'} rescattering, $S_{\rm enh}$, which involves intermediate partons. Since we have to multiply the probabilities of absorption on each individual intermediate parton, the final effect is {\it enhanced} by the large multiplicity of intermediate partons. Unlike $S^2_{\rm eik}({\mathbf b})$, the enhanced survival factor $S^2_{\rm enh}({\mathbf b})$ cannot be considered simply as an overall multiplicative factor. The probability of interaction with a given intermediate parton depends on its position in configuration space; that is, on its impact parameter ${\mathbf b}$ and its momentum $k_t$. This means that $S_{\rm enh}$ simultaneously changes the distribution of the active partons which finally participate in the hard subprocess. It breaks the soft-hard factorization of (\ref{eq:M}).

Do we anticipate that $S_{\rm enh}$ will be important? Working at LO (of the collinear approximation) we would expect that effect may be neglected. Due to strong $k_t$-ordering the
transverse momenta of all the intermediate partons are very large
(i.e. the transverse size of the Pomeron is very small) and therefore the absorptive
effects are negligible. Nevertheless, this may be not true at a very low
$x$, say $x \sim 10^{-6}$, where the parton densities become close to saturation and the
small value of the absorptive cross section is compensated by the large
value of the parton density.
Indeed, the contribution of the first {\it enhanced} diagram, which
describes the absorption of an intermediate parton, was estimated in
the framework of the perturbative QCD in Ref.\cite{BBKM}. It turns out
that it could be quite large. On the other hand, such an effect does not reveal itself
experimentally. The absorptive corrections due to enhanced
screening must increase with energy. This is not observed in the
present data (see \cite{JHEP} for a more detailed discussion).
One reason is that the gap survival factor $S^2_{\rm eik}$
already absorbs almost the whole contribution
from the centre of the disk. The parton essentially only survives eikonal
rescattering on the periphery; that is, at large impact parameters $b$. On the other hand,
on the periphery, the parton density is rather small and the probability
of {\it enhanced} absorption is not large. This fact can be seen in
Ref. \cite{Watt}. There, the momentum, $Q_s$, below which we may approach
saturation, was extracted from the HERA data in the framework of the
dipole model. ($Q_s$ is the inverse size of the dipole for which absorptive corrections become sizeable.) Already at $b=0.6$ fm the value of $Q^2_s$ is such that $Q^2_s<0.3$ GeV$^2$
for the relevant values of $x$, namely $x\sim 10^{-6}$. However, in the perturbative QCD calculations of Refs.~\cite{KMRpr,KMRSchi,KKMRcentr} the infrared cutoff $Q_t>Q_0=0.85$ GeV was introduced in order not to have uncontrollable uncertainties in the parton distribution functions coming from the non-perturbative domain.

Now, the model of Ref.~\cite{KMRnns}, with its multi-component Pomeron, allows us to calculate the survival probability of the rapidity gaps, to {\it both} eikonal and enhanced rescattering. Recall that the evolution equations in rapidity (like (\ref{eq:evol1})) have a matrix form in $aa'$ space, where $a=1,2,3$ correspond to the large-, intermediate- and small-size components of the Pomeron.    We start the evolution from the large component $P_1$, and since the evolution equations allow for a transition from one component to another (corresponding to BFKL diffusion \cite{Lip} in ln$k_t$ space), we determine how the enhanced absorption will affect the high-$k_t$ distribution in the small-size component $P_3$, which contains the active gluon involved in forming the Higgs.
Moreover, at each step of the evolution the equations include absorptive factors of the form $e^{-\lambda(\Omega^a_k +\Omega^a_i)/2}$. By solving the equations with and without these suppression factors, we could quantify the effect of enhanced absorption. However, there are some subtle issues here, see \cite{RMK2}. First, since we no longer have soft-hard factorization, we must first specify exactly what is included in the bare hard amplitude.

Another relevant observation is that the phenomenologically determined generalised gluon distributions are usually taken at $p_t=0$, and then the observed ``total'' cross section is calculated by integrating over $p_t$ of the recoil protons {\it assuming} an exponential behaviour $e^{-Bp_t^2}$; that is
\be
\sigma ~=~\int\frac{d\sigma}{dp_{1t}^2 dp_{2t}^2}dp_{1t}^2 dp_{2t}^2 ~=~\frac{1}{B^2}\left.\frac{d\sigma}{dp_{1t}^2 dp_{2t}^2}\right|_{p_{1t}=p_{2t}=0}~,
\ee
where
\be
\int dp^2_t~e^{-Bp_t^2}~=~1/B~=~\langle p_t^2\rangle.
\ee
However, the total soft absorptive effect changes the $p_t$ distribution in comparison to that for the bare cross section determined from perturbative QCD. Moreover, the correct $p_t$ dependence of the matrix element ${\cal M}$ of the hard $gg \to H$ subprocess does {\it not} have an exponential form. Thus the additional factor introduced by the soft interactions is not just the gap survival $S^2$, but rather $S^2\langle p^2_t \rangle^2$, where the square arises since we have to integrate over the $p_t$ distributions of {\it two} outgoing protons. Indeed in all the previous calculations the soft prefactor had the form\footnote{At larger impact parameter $b$ the absorption is weaker. Hence the value of $S^2$ increases with the slope $B$. It was shown that the ratio $S^2/B^2$ is approximately stable for reasonable variations of $B$ \cite{KMRSchi}.} $S^2/B^2$.  Note that, using the model of Ref.~\cite{KMRnns}, we no longer have to {\it assume} an exponential ${\mathbf b}$ behav!
 iour of the matrix element. Now the ${\mathbf b}$ dependence of ${\cal M}({\mathbf b})$ is driven by the opacities, and so is known. Thus we present the final result in the form $S^2\langle p^2_t \rangle^2$. That is, we replace $S^2/B^2$ in (\ref{eq:M}) by $S^2\langle p^2_t \rangle^2$.  So if we wish to compare the improved treatment with previous predictions obtained assuming $B=4~{\rm GeV}^{-2}$ we need to introduce the ``renormalisation'' factor $(\langle p_t^2 \rangle B)^2$. The resulting (effective) value is denoted by $S^2_{\rm eff}$.

\vspace{0.5cm}
Before we do this, there is yet another effect that we must include. We have to allow for a threshold in rapidity
\cite{thr,JHEP,kkmr}.
 The evolution equation for $\Omega^a_k$, (\ref{eq:evol1}), and the analogous one for $\Omega^a_i$, are written in the leading ln$(1/x)$ approximation, without any rapidity threshold. The emitted parton, and correspondingly the next rescattering, is allowed to occur just after the previous step. On the other hand, it is known that a pure kinematical $t_{\rm min}$ effect suppresses the probability to produce two partons close to each other. Moreover, this $t_{\rm min}$ effect becomes especially important near the production vertex of the heavy object. It is, therefore, reasonable to introduce some threshold rapidity gap, $\Delta y$, and to compute $S^2_{\rm enh}$ only allowing for absorption outside this threshold interval, as indicated in Fig.~\ref{fig:pAp}. For exclusive Higgs boson production at the LHC\footnote{A very small value $S^2_{\rm enh}=0.063$ is claimed in
\cite{GLMM}, which would translate into an {\it extremely} small value of
$S^2_{\rm eff}=0.0235 \times 0.063=0.0015$. There are many reasons why this estimate is invalid.
In this model the two-particle irreducible amplitude
depends on the impact parameter $b$ {\it only} through the form factors of the incoming
protons. The enhanced absorptive effects (which result
from the sum of the enhanced diagrams) are the same
at any value of $b$. Therefore, the enhanced screening effect does not
depend on the initial parton density at a particular impact parameter $b$,
and does not account for the fact that at the periphery of the proton, from where the main
contribution comes (after the $S_{\rm eik}$ suppression), the parton density is
much smaller than that in the centre. For
this reason the claimed value of $S^2_{\rm enh}$ is much too small. Besides
this lack of $k_t \leftrightarrow b$ correlation, the model has no diagrams
with odd powers of $g_{3P}$. For example, the lowest triple-Pomeron diagram is
{\it missing}. That is, the approach does not contain the first, and most important at the lower energies, absorptive correction. Next, recall that in a theory which contains the
triple-Pomeron coupling only, without the four-Pomeron term (and/or 
more complicated multi-Pomeron vertices), the total cross section
{\it decreases} at high energies \cite{GK}. On the other hand, the
approximation used in \cite{GLMM} leads to 
saturation (that is, to a constant cross section) at very high energies. In other words, the approach is not valid at high energies. This means that such an approximation can only be justified in a limited energy interval; 
at very high energies it is inconsistent with asymptotics, while
at relatively low energies the first term, proportional to the first power of the
triple-Pomeron coupling $g_{3P}$, is missing.
Finally, the predictions of the model of \cite{GLMM} have not been compared to the CDF
exclusive data of Section \ref{sec:cdf}.}, the model gives $S^2_{\rm eff}=0.004,\; 0.009$ and 0.015 
for $\Delta y=0,\; 1.5$ and 2.3 respectively \cite{RMK2}. For $\Delta y=2.3$ all the NLL BFKL
corrections \cite{BFKLnnl} may be reproduced by the threshold
effect \cite{thr,JHEP,kkmr,salam}.  

\vspace{0.2cm}
Furthermore, Ref.~\cite{RMK2} presents arguments that
\be
\langle S^2_{\rm eff} \rangle~=~0.015^{~+0.01}_{~-0.005}
\label{eq:s}
\ee
should be regarded as a {\it conservative} (lower) limit for the gap survival probability in the exclusive production of a SM Higgs boson of mass 120 GeV at the LHC energy of $\sqrt{s}=14$ TeV.  Recall that this effective value should be compared with $S^2$ obtained using the exponential slope $B=4~{\rm GeV}^{-2}$. The resulting value for the cross section is, conservatively,
\be
\sigma(pp \to p+H+p)~\simeq~2-3 ~{\rm fb},
\label{eq:x}
\ee
with an uncertainty\footnote{Besides the uncertainty arising from that on $ S^2_{\rm eff}$, the other main contribution to the error comes from that on the unintegrated gluon distributions, $f_g$, which enter to the fourth power \cite{KMRearly,clp}.}
of a factor of 3 up or down, see also \cite{KKMRext,KMRearly}. In Section \ref{sec:cons} we explain why the values (\ref{eq:s}) and  (\ref{eq:x}) should be regarded as conservative.

\subsection{Exclusive processes observed at the Tevatron \label{sec:cdf}}
\begin{figure}
\begin{center}
\includegraphics[height=3cm]{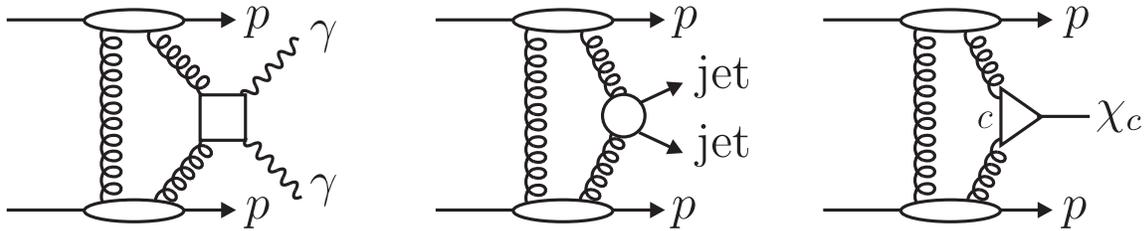}
\caption{The mechanism for the exclusive processes observed by CDF at the Tevatron. The survival probabilities of the rapidity gaps are not shown in the sketches.  }
\label{fig:cdf}
\end{center}
\end{figure}  
Exclusive diffractive processes of the type $\bar{p}p \to \bar{p}+A+p$ have {\it already} been observed by CDF at the Tevatron, where $A=\gamma\gamma$ \cite{CDFgg} or dijet \cite{CDFdijet} or $\chi_c$ \cite{CDFchi}.   As the sketches in Fig.~\ref{fig:cdf} show, these processes are driven by the same mechanism as that for exclusive Higgs production, but have much larger cross sections. They therefore serve as ``standard candles''.

CDF observe three candidate events for $\bar{p}p \to \bar{p}+\gamma\gamma+p$ with $E_T^{\gamma}>5$ GeV and $|\eta^{\gamma}|<1$ \cite{CDFgg}. Two events clearly favour the $\gamma\gamma$ hypothesis and the third is likely to be of $\pi^0 \pi^0$ origin. The predicted number of events for these experimental cuts is $0.8^{+1.6}_{-0.5}$ \cite{KMRSgg}, giving support to the theoretical approach used for the calculation of the cross sections for exclusive processes. In fact, central exclusive two- and three-jet production offers more detailed tests of the theoretical formalism, see \cite{KMRearly}.

Especially important are the recent CDF data \cite{CDFdijet} on
 exclusive production
of a pair of high $E_T$ jets, $p\bar {p} \to p+jj+\bar {p}$.
As discussed in \cite{KMRjj,KMR,KMRpr} such measurements could provide
 an effective $gg^{PP}$ `luminosity
\begin{figure} [h]
\begin{center}
\includegraphics[height=7.5cm]{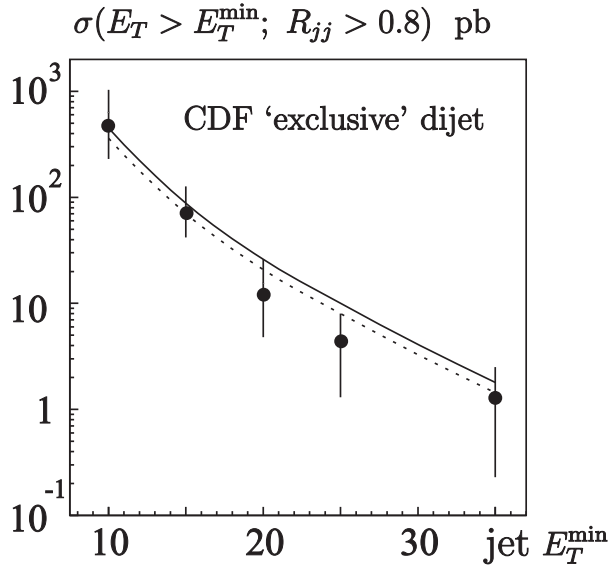}
\caption{The cross section  for `exclusive' dijet production
 at the Tevatron as a function $E_T^{\rm min}$ as measured by CDF \cite{CDFdijet}. 
  The data integrated over
the domain $R_{jj} \equiv M_{\rm dijet}/M_{PP} > 0.8$ and $E_T > E_T^{\rm min}$. A jet cone of $R<0.7$ is used.
The curves are the pure exclusive cross section
calculated \cite{KMRjj,KMRpr} using the CDF event selection. 
The solid curve is obtained \cite{vakjj} by rescaling the parton (gluon)  transverse momentum
$p_T$ to the measured jet
transverse energy $E_T$ by  $E_T=0.8 p_T$. The dashed curve assumes
$E_T=0.75 p_T$. The rescaling procedure effectively accounts for the hadronization
and radiative 
effects, and for jet energy losses outside the selected jet cone.}
\label{fig:sjj}
\end{center}
\end{figure}
monitor' just in the kinematical region appropriate for Higgs
production. The corresponding cross section was evaluated to
be about 10$^4$ times larger than that for the production of a SM Higgs boson.
Since the exclusive dijet cross section is rather large, this  process appears
to be an ideal `standard candle'.
A comparison of the data with 
analytical predictions \cite{KMRjj,KMRpr}  is given in Fig.~$\ref{fig:sjj}$. 
It shows the  $E_T^{\rm min}$ dependence  
for the dijet events with $R_{jj} \equiv M_{{\rm dijet}}/M_{PP} > 0.8$,
where $M_{PP}$ is the invariant energy of the incoming
Pomeron-Pomeron system.  The agreement with the theoretical expectations
\cite{KMR,KMRpr} lends credence to the predictions for the exclusive
Higgs production. 

In particular, these CDF dijet data clearly demonstrate the suppression due to the Sudakov factor $T$ which occurs in the evaluation of the unintegrated gluon distribution, $f_g$, needed to predict the exclusive cross section, (\ref{eq:M}). On dimensional grounds we would expect $d\sigma/dE_T^2 \propto 1/E_T^4$. This behaviour is modified by the anomalous dimension of the gluon and by a stronger Sudakov suppression with increasing $E_T$. Already the existing CDF exclusive dijet data \cite{CDFdijet} exclude predictions which omit the Sudakov effect. In Section \ref{sec:jjj} we describe how observation of central exclusive {\it three-jet}, as well as two-jet, production at the LHC  offers more detailed tests of $f_g$.

CDF \cite{CDFchi} have also recently observed exclusive $\chi_c$ production, $p\bar {p} \to p+\chi_c+\bar {p}$, via the decay chain $\chi_c \to J/\psi+\gamma \to \mu^+\mu^-\gamma$. They collected $65 \pm 10$ signal events, with limited $M(J/\psi\gamma)$ resolution. Assuming the dominance of $\chi_c(0)$ production, this corresponds to 
\be
d\sigma(\chi_c)/dy|_{y=0}~=~76 \pm 14 ~{\rm nb}.
\label{eq:chi}
\ee
Strictly speaking, due to the low scale associated with exclusive $\chi_c$ production, only order-of-magnitude estimates can be made for the cross section. Ref.~\cite{KMRSchi} assumes a perturbative contribution coming from integrating round the gluon loop in Fig.~\ref{fig:pAp} for $Q_t>0.85$ GeV, with the remaining imfrared contribution estimated non-perturbatively. This gave a prediction for $d\sigma(\chi_c(0^{++})/dy|_{y=0}$ of 90 nb,\footnote{The value 130 nb obtained in \cite{KMRSchi} has been changed to 90 nb to take into account the revised value of the total $\chi_c(0)$ width of 10.2 MeV as compared to the value in the RPP(2002) which was 1.45 times higher.}
 with $\frac{1}{3}$ coming from the non-perturbative region and $\frac{2}{3}$ from the perturbative domain. The rapidity gap survival factor was taken to be $\langle S^2 \rangle =0.07$, assuming only eikonal rescattering. However, for the relatively light $\chi_c(0)$, the available rapidity interval for the enhanced suppression is large. In this case the inclusion of enhanced rescattering may cause a large difference -- reducing the exclusive cross section by up to $\frac{1}{3}$. The resulting cross section of 30 nb, which includes $S_{\rm enh}$, is below the value given in (\ref{eq:chi}).

However, in the conditions of the experiment \cite{CDFchi}, where the transverse momenta of the forward outgoing protons are not measured, the situation for $\chi_c$ may be more complicated, due to possible contributions from the higher spin $\chi_c(1^{++})$ and $\chi_c(2^{++})$ states. Recall that 65 events with two large rapidity gaps were observed via the $\chi_c \to J/\psi\gamma$ decay mode, and that the $\chi_c(0)$ branching fraction to this mode is very low. Explicitly, the $\chi_c(0),~\chi_c(1),~\chi_c(2)$ branching fractions to $J/\psi\gamma$ are 0.013, 0.36 and 0.20 respectively. Also, the bare cross sections for $\chi_c(1)$ and $\chi_c(2)$ production are suppressed relative to $\chi_c(0)$ production by factors of $\langle p^2_t \rangle /M^2$ and ${\langle p^2_t \rangle}^2 /Q_t^4$ respectively, where $p_t$ is the transverse momentum of the outgoing protons and $Q_t$ is that round the gluon loop. Numerically this leads to a suppression by about a factor of $30-100$, depending on the choices of the global partons and of the infrared cutoff that are used in the computation. Part of this suppression is compensated by a larger gap survival probability for $\chi_c(1)$ and $\chi_c(2)$, since, due to their spin structure, they are produced more peripherally. Thus, finally, we estimate a production ratio $\chi_c(0)/\chi_c(1) \simeq 10-40$, with more or less the same suppression ratio found for $\chi_c(2)$\footnote{Recall that these predictions at such low scales have a large uncertainty. Moreover, the NLO corrections could be quite sizeable, and require further detailed studies. Note, also, that the relative number of events where the forward protons dissociate is larger for $\chi_c(1)$ and $\chi_c(2)$ than for $\chi_c(0)$.}. In summary, since the energy resolution of the CDF observation is insufficient to separate the three $\chi_c$ states, then, allowing for the $J/\psi\gamma$ branching fractions, we may expect more or less comparable contributions from each of the states\footnote{The importance of $\chi_c(1)$ production has been recently emphasized in Ref.~\cite{pst}.}. Therefore even the conservative expectation of 30 nb for $\chi_c(0)$ is not inconsistent with (\ref{eq:chi}). 

To make further progress, it would be instructive to observe central exclusive $\chi_c$ production via the $\pi\pi$ or $K\bar{K}$ decay channels, see \cite{KMRSchi}. Recall that the $\pi\pi$ or $K\bar{K}$ decay modes of $\chi_c(0)$ have a branching fraction of about 1$\%$, while these decay channels are forbidden for $\chi_c(1)$, and suppressed by about a factor of 5 for $\chi_c(2)$ relative to $\chi_c(0)$; rather than enhanced relative to $\chi_c(0)$, as was the case for the $J/\psi\gamma$ channel. Alternatively we could study the angular distribution of the $J/\psi\gamma$ system in order to distinguish the spins of the parent $\chi_c$.

As a final general comment, we note that it is sometimes stated that enhanced diagrams, which break soft-hard factorization, generate an extremely small gap survival factor, $S_{\rm enh}$ \cite{GLMM,strik}. However,
the analysis of \cite{RMK2} shows the following hierarchy of the size of the  gap survival factor to enhanced rescattering 
\be
S^{\rm LHC}_{\rm enh}(M_H>120 ~{\rm GeV})~~>~~S^{\rm Tevatron}_{\rm enh}(\gamma\gamma;E_T>5~{\rm GeV})~~>~~S^{\rm Tevatron}_{\rm enh}(\chi_c),
\ee
which reflects the size of the various rapidity gaps of the different exclusive processes. The fact that $\gamma\gamma$ and $\chi_c$ events have been observed at the Tevatron confirms that there is no danger that enhanced absorption will strongly reduce the exclusive SM Higgs signal at the LHC energy.

\subsection{Dependence of the survival factors on collider energy}

It is relevant to ask how the survival factors, $S^2_{\rm eik}$ and $S^2_{\rm enh}$, change in size in going from the Tevatron to the LHC energy. The energy dependence of the eikonal gap survival $S^2_{\rm eik}$
is driven by the ratio $\sigma_{\rm tot}/B_{\rm el}$, which controls the behaviour of the opacity $\Omega$. Here, $B_{\rm el}$ is the slope of the elastic differential cross section. As we have seen, the growth of the total cross section is not expected to be large, and is partly compensated by the increase in the value of the slope $B_{\rm el}$.
Indeed, for exact ``geometric scaling'', for which $\sigma_{\rm tot}\propto R^2$ (where $R$ is the interaction radius), the function $\Omega(b)$ does not depend on energy at all.
Thus we expect a rather weak (only logarithmic) decrease of the eikonal survival probability with energy.

The enhanced survival factor $S^2_{\rm enh}$ may decrease with energy due to 
\begin{itemize} 
\item [(i)] the increased size of the rapidity interval ($\propto {\rm ln}{(s/M^2})$) available for screening corrections,
\item [(ii)] the increase in the parton density as $x \sim 1/s$ decreases,
\end{itemize}
where we are concerned with the part of the amplitude denoted by $S_{\rm enh}$ in Fig. 8. We discuss these possible $s$ dependences in turn.

At fixed $M^2$, the available rapidity interval grows with $s$ and this should lead to a decrease of $S^2_{\rm enh}$. Recall that $M$ is the mass of the centrally produced system. However, no sizeable variation with energy is observed in present data, say, in leading neutron production at HERA. This phenomenological fact indicates that the role of $S^2_{\rm enh}$ factor should be relatively small (see \cite{JHEP} for a detailed discussion).

The values of $x\sim 10^{-6}$ relevant at LHC energies are quite small. At leading order (LO) the gluon density increases with $1/x$. So, at first sight, it appears that this may lead to the Black Disk Regime where the enhanced absorptive corrections will cause the cross section for exclusive production to practically vanish. However, NLO global parton analyses show that at relatively low scales ($k^2_t=\mu^2\sim 2\ -\ 4$ GeV$^2$) the gluons are flat for $x<10^{-3}\, -\, 10^{-4}$ (or even decrease when $x\to 0$).
Recall that the contribution of enhanced diagrams from  larger scales
decrease as $1/k^2_t$ (see \cite{BBKM}). The anomalous dimension
$\gamma <1/2$ is not large and so the growth of the gluon density, $xg(x,\mu^2)\propto (\mu^2)^\gamma$, cannot compensate the factor $1/k^2_t$.
Therefore the whole enhanced diagram contribution should be evaluated at rather low scales where the NLO gluon is flat\footnote{Note that the flat $x$-behaviour of NLO gluons allow the justification of the inequalities in (20).}
 in $x$. Thus there is no reason to expect a strong energy dependence of $S^2_{\rm enh}$.

\subsection{Why the estimates are conservative \label{sec:cons}}

It is useful to list the reasons why we regard the estimates of the rapidity gap survival factor (\ref{eq:s}) and the exclusive
cross section (\ref{eq:x}) as conservative lower limits.

First, we should not use the whole irreducible amplitude $\Omega^{\rm tot}_{ik}({\mathbf b})$ when calculating the absorption (as, for example, in exp$(-\Omega^{\rm tot}_{ik}({\mathbf b})/2)$ of (\ref{eq:c3pp})), but only that part of it which corresponds to having some rescattering which populates the gap in the chosen rapidity interval. The computations of \cite{KMRnns,RMK2} neglect this effect. This means that the gap survival probabilities, and the true cross sections of diffractive dissocation, should be a bit larger than the predictions given earlier. 

Second, a detailed study \cite{RMK2} of the $b$ dependence of $S^2_{\rm enh}$, obtained from the soft model of \cite{KMRnns}, is instructive. Assuming that all the partons are distributed {\it homogeneously}, we would expect $S^2_{\rm enh} \to 1$ at the (large) values of $b$ occurring in the periphery of the $pp$ interaction. However, the results show only a weak tendency to increase at large $b$. This may be explained by the fact that in the `soft' model of \cite{KMRnns} (with its small value of $\alpha'_P$), $S^2_{\rm enh}$ comes mainly from ``{\it hot-spots}'' in which many individual intermediate partons are concentrated within small $b$ domains. In other words, most of enhanced absorption occurs within the same parton shower, and is due to secondaries produced during the evolution with practically the {\it same} impact parameter $b$. On the other hand, a detailed analysis \cite{Watt} of HERA data shows that the value of the saturation scale $Q_s(b)$ decreases rapidly with increasing $b$. (Recall that $Q_s$ is the inverse size of the dipole for which absorptive corrections become important.) This indicates that the parton-parton correlations are too strong in the present model. Therefore, we consider our results are close to the maximum possible gap suppression.

Third, the partonic nature of the diffractive eigenstates is not specified in the soft model of \cite{KMRnns}. When BFKL is introduced, it is implied that the partons are gluons, and so the screening exponents were chosen assuming gluons in both Pomerons. However, at  NLO the gluon density in the relevant domain
is approximately flat in $x$. In terms of QCD the increase of the parton
density with decreasing $x$, at the relevant low scales, may reflect the growth of the NLO sea quark
density with decreasing $x$. On the other hand, the central exclusive
amplitudes of interest are driven by gluon-gluon
fusion. If there is  screening it should be caused by the
(NLO) sea-quark contribution in these
amplitudes. Then the effective colour factor is smaller
than in the model of \cite{KMRnns} where BFKL (that is, pure gluonic colour
coefficients) is used to evaluate the absorptive effects.
Hence, again, we expect a {\it larger} value of $S_{\rm enh}$.

Finally, it is worth mentioning that in deriving predictions for the cross sections of central exclusive processes, such as (\ref{eq:x}), rather conservative choices of parton distributions were used \cite{KMRpr,KKMRext,KMRSchi}. Larger predictions will result if the gluon densities are found to be even a little larger, recall the $(xg)^4$ dependence.

We note from Section \ref{sec:cdf} that, on average, the predictions lie a little below the present exclusive data obtained by CDF at the Tevatron. However this is in line with the above comments. It will be informative to see how the data/theory comparisons hold up with the improved experimental statistics which will come from the Tevatron and, later on, from the LHC.

\subsection{Early LHC probes of exclusive production  \label{sec:jjj}}
Recall that the uncertainties associated with predictions for an exclusive process are potentially not small. Each stage has its own uncertainties. Therefore, it is important to perform checks of the approach using processes with appreciable cross sections that will be experimentally accessible in the first data runs of the LHC; that is with integrated luminosities in the range 100 ${\rm pb}^{-1}$ to 1 ${\rm fb}^{-1}$. In fact, it is possible identify processes where the different ingredients of the formalism used to calculate central exclusive production can be tested experimentally, more or less independently \cite{KMRearly}.

To probe the gap survival factor to eikonal rescattering, $S^2_{\rm  eik}$,  we need  a process with 
 a bare cross section that can be calculated reliably.
 Good candidates are the production of $W$ or $Z$ bosons with rapidity gaps on either side \cite{KMRearly}.
\begin{figure}
\begin{center}
\includegraphics[height=5cm]{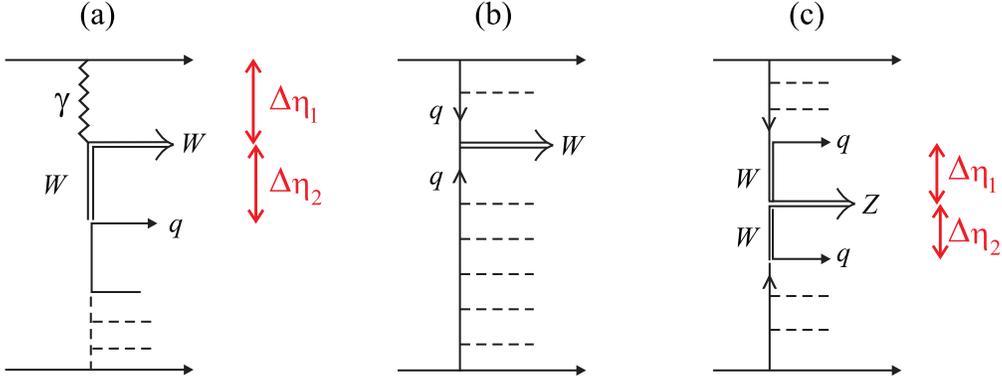}
\caption{(a) $W$ production with 2 gaps, (b) Inclusive $W$ production, (c) $Z$ production with 2 gaps.}
\label{fig:WZ}
\end{center}
\end{figure}
In the case of `$W$+gaps' production the main contribution comes from the diagram of
 Fig.~\ref{fig:WZ}(a). One gap, $\Delta \eta_1$, is associated with photon exchange, 
while the other, $\Delta \eta_2$, is associated with $W$ exchange.

At first sight, the probability for soft rescattering in such a process is rather small and we would expect $S^2 \sim 1$. The reason is that the transverse momentum, $k_t$, distribution of the exchanged photon is given by the logarithmic integral
\be
\int \frac{dk^2_t~k_t^2}{(|t_{\rm min}|+k_t^2)^2}~,~~~~~~{\rm with}~~~|t_{\rm min}|~\simeq~\frac{m_N^2 \xi^2}{1-\xi}~,
\label{eq:tmin}
\ee
for which the dominant contribution comes from the low $k_t^2$ region. In other words, the main contribution comes from the region of large impact parameters, $b$, where the opacity of the proton is small. However, the minimum value of $|t|$,
is not negligibly small. Note that the momentum fraction $x_p=1-\xi$ associated with the upper proton can be measured with sufficient accuracy\footnote{The CDF collaboration \cite{cdf} have demonstrated that this method provides an accurate determination of $\xi$.}, {\it even without the tagging of the forward protons}, by summing the momentum fractions
\be
\xi_i~=~\sqrt{m_i^2+k_{ti}^2}~e^{y_i}/\sqrt{s}
\label{eq:xi}
\ee
of the outgoing $W$ and the hadrons observed in the calorimeters.  As long as the gap $\Delta \eta_2$ is large, the dominant contribution to the sum $\xi=\sum \xi_i$ comes from the decay products of the $W$ boson. For example, for $\eta_W=2.3(-2.3)$, we expect an $\xi$ distribution centred about $\xi \sim 0.1(0.001)$.
If we take $\Delta \eta_2 >3$, then the cross section $d\sigma(W+{\rm gaps})/d{\rm ln}\xi$ is $0.1(1)$ pb and $S^2_{\rm eik} \sim 0.87(0.45)$ for $\xi \sim 0.001(0.1)$ and $\eta_W=2.3(-2.3)$, respectively.

In the first LHC data runs it may be difficult to measure the absolute value of the cross section with sufficient accuracy. Most probably the ratio ($W$+gaps/$W$ inclusive) will be measured first. In this case, the inclusive $W$ production process (in the same kinematic region, Fig.~\ref{fig:WZ}(b)) plays the role of the luminosity monitor. Note that the cross section for inclusive $W$ production is much larger than that with rapidity gaps. The reason is that an inclusive $W$ is produced directly by $q\bar{q}$ fusion, which is prohibited for gap events since the colour flow produced by the $t$-channel quarks populates the $\Delta \eta_{1,2}$ rapidity gaps. 

Of course, the survival factor $S^2$ measured in $W$+gaps events is quite different from that for exclusive Higgs production, which comes from smaller values of $b$. Nevertheless this measurement is a useful check of the model for soft rescattering.
A good way 
to study the low impact parameter, $b$, region 
 is to observe $Z$ boson production via $WW$ fusion, see Fig.~\ref{fig:WZ}(c). Here, both gaps
 originate from $W$-exchange, and the corresponding $b$ region 
 is similar to that for exclusive Higgs production. 
 The expected $Z$+gaps cross section is of the order of 0.2 pb,
 and $S^2$=0.3 for $\Delta \eta_{1,2} > 3$ and for quark jets with $E_T>50$ GeV \cite{krsw}.

How can the LHC probe the unintegrated gluon $f_g$ distribution? Recall that the exclusive cross section, (\ref{eq:M}), depends on $f_g$ evaluated in terms of the integrated gluon density $g(x,Q_t^2)$
for $Q_t^2 \sim 4~\GeV^2$ 
and $x \sim M_H/\sqrt{s}$. Now for $Q_t^2 = 4~\GeV^2$ and $x=10^{-2}$ the MSTW \cite{MSTW} and CTEQ \cite{CTEQ} global parton analyses indicate that $xg$ lies in the interval (3.2, 3.8).
This is a big uncertainty bearing in mind that the exclusive cross section 
 depends on $(xg)^4$.
\begin{figure} [t]
\begin{center}
\includegraphics[height=4cm]{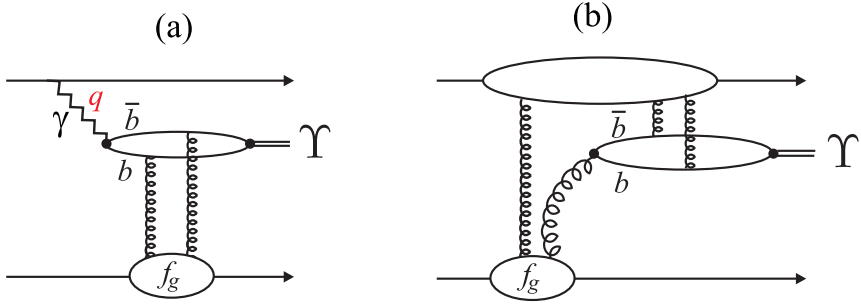}
\caption{Exclusive $\Upsilon$ production via (a) photon exchange, and (b) via odderon exchange.}
\label{fig:upsilon}
\end{center}
\end{figure}
To reduce the uncertainty associated with $f_g$ we can measure exclusive $\Upsilon$ production at the LHC.
 The process is shown in Fig.~\ref{fig:upsilon}(a). 
  The cross section for $\gamma p \to \Upsilon p$ is given in terms
   of the same  unintegrated gluon distribution $f_g$ that occurs in Fig.~\ref{fig:pAp}.
There may be competition between production via photon exchange\footnote{The first experimental data on exclusive $J/\psi$ production obtained by the CDF collaboration \cite{CDFchi} are consistent with the theoretical expectations for photon exchange \cite{bmsc}, see also \cite{KMRphot}.} 
Fig.~\ref{fig:upsilon}(a),
 and via odderon exchange,
 see Fig.~\ref{fig:upsilon}(b).
A lowest-order  calculation (e.g. \cite{bmsc})
 indicates that the odderon process (b) may be comparable to the photon-initiated process (a). 
 If the upper proton is tagged, it will be straightforward to separate the two mechanisms; larger $p_t$ is expected for the odderon process of Fig.~\ref{fig:upsilon}(b).

\begin{figure} [t]
\begin{center}
\includegraphics[height=9.5cm]{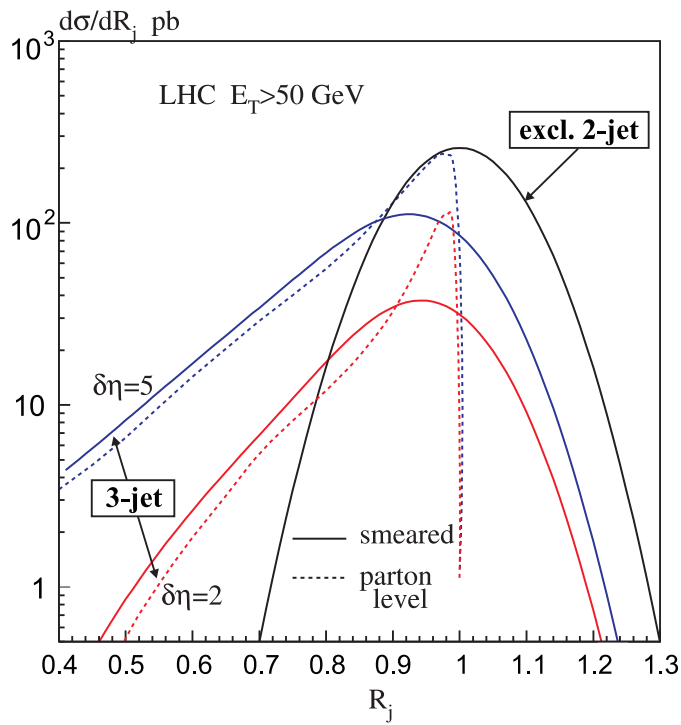}
\caption[The cross section]{The $R_j$ distribution of exclusive two- and three-jet production at the LHC \cite{KMRrj}.
Without smearing, exclusive two-jet production would be just a $\delta$-function at $R_j=1$.
The distribution for three-jet production is shown for two choices of the rapidity interval, $\delta\eta$, containing the jets; these distributions are shown with and without smearing. Here, we have taken the highest $E_T$ jet to have $E_T>50$ GeV. To indicate the effect of jet smearing, we have assumed a Gaussian distribution with a typical
resolution $\sigma=0.6/\sqrt{E_T~{\rm in~ GeV}}$.}
\label{fig:Rj}
\end{center}
\end{figure}
A more detailed probe of $f_g$ may come from studies of central exclusive jet production at the LHC. We have already noted that the computation of $f_g$ in terms of the integrated distribution $g$ involves the presence of a Sudakov-like form factor $T$, which ensures the absence of gluon emission in the evolution from $Q_t$ up to the hard scale $\sim M_A/2$; and that $T$ provides the infrared stability of the $Q_t$ integral in (\ref{eq:M}) for sufficiently large $M_A$.
Also recall from Fig.~\ref{fig:sjj} that a measure of exclusive dijet production at the 
Tevatron, $p\bar{p} \to p+jj+\bar{p}$, was obtained \cite {CDFdijet} by plotting 
the cross section in terms of the variable
$R_{jj}=M_{jj}/M_A$, where $M_A$ is the mass of the whole central system.
However, 
the $R_{jj}$  distribution is smeared out by
QCD radiation, hadronization, the jet algorithm and 
 other experimental effects \cite {CDFdijet,KMRrj}.
To weaken the smearing it was proposed in Ref.~\cite{KMRrj} 
to study the  dijets  in terms of the  variable
$R_j~=~2E_T ~({\rm cosh}~\eta^*)/M_A~$,
where only the transverse energy and the rapidity $\eta$ of the jet with
the {\it largest} $E_T$ enter.
 Here $\eta^* = \eta_{\rm jet} -y_A$, where $y_A$ is the rapidity of the  central system.
Clearly,
the largest $E_T$ jet is less affected by the smearing.  

In Fig.~\ref{fig:Rj} we show the $R_j$ distribution of both exclusive two- and three-jet production expected at the LHC. For three-jet production we show predictions for two choices of the rapidity interval $\delta\eta$ within which all three jets must lie. If we take the largest $E_T$ jet to have $E_T>50$ GeV at the LHC, we see that the cross section for exclusive three-jet production reaches a value of the order of 100 pb. Of course, if we enlarge the rapidity interval $\delta\eta$ where we allow emission of the third jet, then $d\sigma/dR_j$ will increase, see Fig.~\ref{fig:Rj}. Indeed, the measurement of the exclusive two- and three-jet cross sections {\it as a function of $E_T$} of the highest jet allows a detailed check of $f_g$ and the Sudakov factor $T$; with much more information coming from the observation of the $\delta\eta$ dependence of three-jet production.  Note that the background from the inclusive interaction of two soft Pomerons (so-called double-Pomeron-e!
 xchange) should be small for $R_j \gapproxeq 0.5$, and can be removed entirely by imposing an $E_T$ cut on the third jet, say $E_T>5$ GeV.

Finally, how can the LHC probe the gap survival probability, $S^2_{\rm enh}$, to enhanced rescattering, which violates soft-hard factorization? It appears at first sight, that we may study the role of enhanced absorption by observing the $W$+2 gaps process shown in Fig.~\ref{fig:WZ}(a). To do this one may vary the transverse momentum of the accompanying quark jet, $q$, and the size of the rapidity gap $\Delta \eta_2$. For lower transverse momentum of the quark jet we expect a stronger absorptive effect, that should decrease with increasing $\Delta \eta_2$, since the number of partons in the rest of the rapidity interval increases. Unfortunately, the photon-initiated process of Fig.~\ref{fig:WZ}(a) occurs at large impact parameter $b$ where the probability of rescattering is small. Moreover it will challenging to observe the quark jet at low $E_T$ in the relevant rapidity interval $(-5<\eta<-3)$. Therefore we will discuss other processes which depend on rescattering on intermediate pa!
 rtons. 

\begin{figure}[ht]
\begin{minipage}[b]{0.45\linewidth}
\centering
\includegraphics[height=3.8cm]{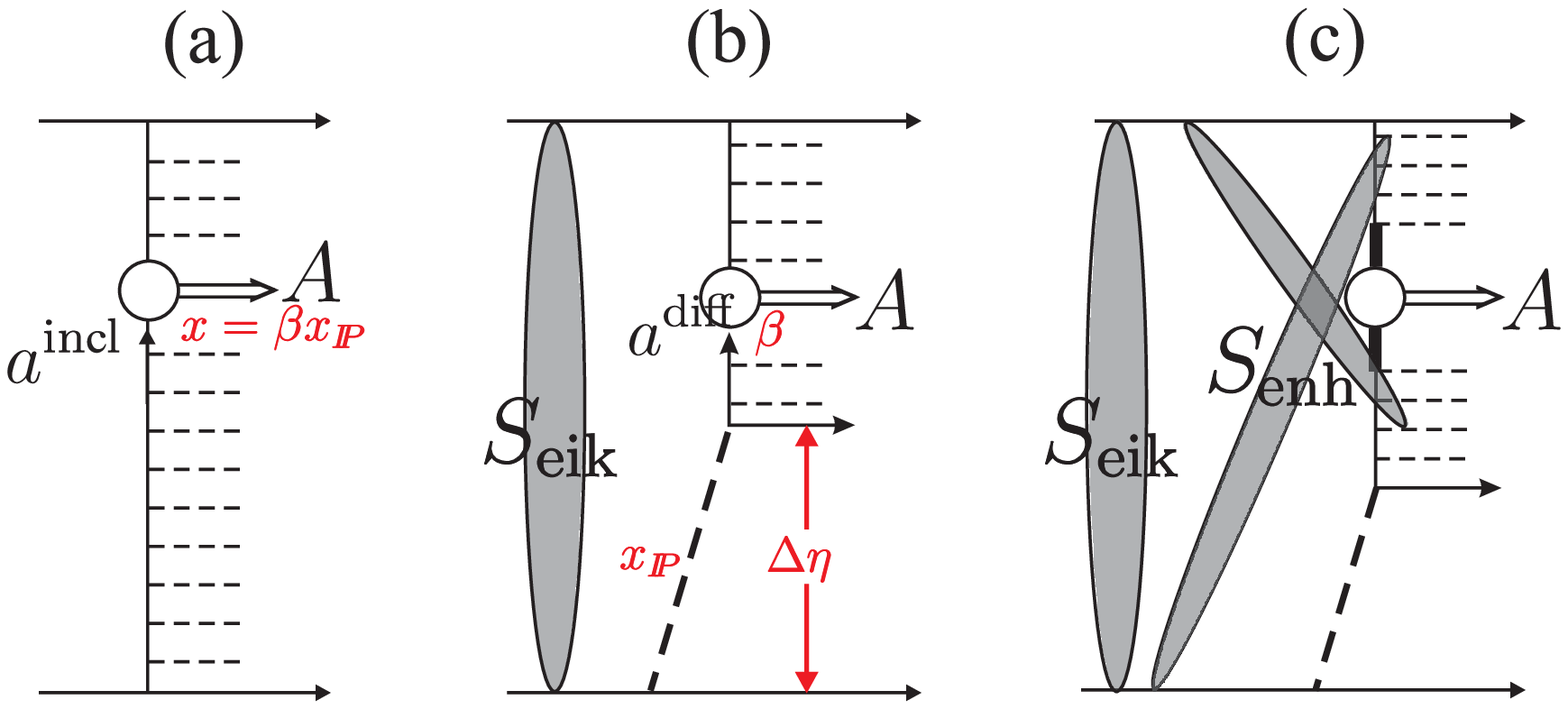}
\caption{Schematic diagrams for (a) the inclusive production of a system $A$, (b) and (c) for the diffractive production of $A$ without and with `enhanced' soft rescattering on intermediate partons. The system $A$ is taken to be either a $W$ boson or an $\Upsilon$ or a pair of high $E_T$ jets.}
\label{fig:3en}
\end{minipage}
\hspace{0.5cm}
\begin{minipage}[b]{0.5\linewidth}
\centering
\includegraphics[height=8.5cm]{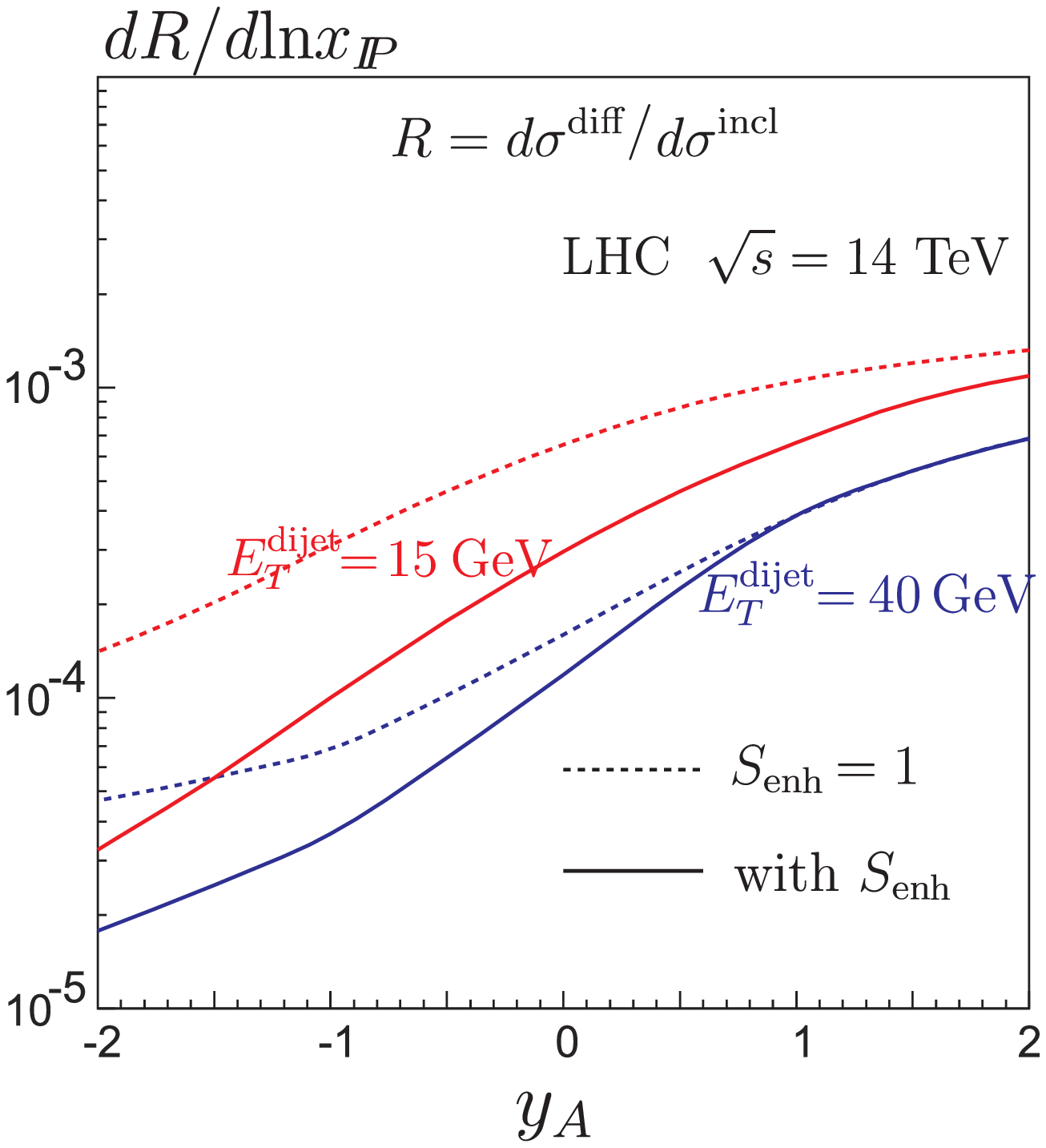}
\caption[The predictions]{The predictions of the ratio $R$ of (\ref{eq:Ren})
 for the production of a pair of high $E_T$ jets.} 
\label{fig:upsd}
\end{minipage}
\end{figure}
The observations we have in mind are the measurements of ratio $R$ of diffractive (one-gap) events for $W$ (or $\Upsilon$ or dijet) production as compared to the number of events for the inclusive process (shown in Fig.~\ref{fig:WZ}(b) for $W$ production). These processes are shown schematically in Fig.~\ref{fig:3en}. In other words, $R$ is the ratio of the process in diagram (c) to that in diagram (a). That is
\be 
R~~=~~\frac{{\rm no.~of}~ (A+{\rm gap)~ events}}{{\rm no.~of~ (inclusive}~A)~{\rm events}}~~=~~
\frac{a^{\rm diff}(x_\funp ,\beta,\mu^2)}{a^{\rm incl}(x=\beta x_\funp,\mu^2)}~\langle S^2_{\rm eik}S^2_{\rm enh}\rangle_{{\rm over}~b},
\label{eq:Ren}
\ee
where $a^{\rm incl}$ and $a^{\rm diff}$ are the parton densities known from the global analyses of inclusive and diffractive deep inelastic scattering data, respectively. The heavy central system $A$ is either $W$ or a pair of high $E_T$ jets or $\Upsilon$ or a Drell-Yan $\mu^+\mu^-$ pair. For $W$ or $\mu^+\mu^-$ pair production the parton densities $a$ are quark distributions, whereas for dijet or $\Upsilon$ production they are mainly gluon densities.  Thus measurements of the ratio $R$ will probe the gap survival factor averaged over the impact parameter $b$.

To demonstrate the possible size of the effect, a simple model was used to estimate $S^2_{\rm enh}$ \cite{KMRearly}. The results for the dijet case 
are shown by the dashed curves in Fig.~\ref{fig:upsd}
  as a function of the rapidity $y_A$ of the 
dijet system.
The enhanced rescattering reduce the ratios
  and lead to steeper $y_A$ distributions, as illustrated by the continuous curves. Perhaps the most informative probe of $S^2_{\rm enh}$ 
is to observe the ratio $R$ for dijet production in the region $E_T \sim 15-30$ GeV.
 For example, for $E_T \sim$ 15 GeV 
 we expect $S^2_{\rm enh} \sim$ 0.25, 0.4 and 0.8 at $y_A=-2, ~0$ and $2$ respectively.

\section{Conclusions}

We have described a model of soft high-energy $pp$ (and $p{\bar p}$) interactions with multi-components in both the $s$- and $t$-channels. The parameters of the model were adjusted to reproduce all the available data in the CERN-ISR to Tevatron energy range, and predictions are made for the LHC energy $\sqrt{s}$=14 TeV. The inclusion of absorptive effects are vital. Low-mass diffractive dissociation is described in terms of a multichannel eikonal, whereas high-mass dissociation is included using a full set of multi-Pomeron-Pomeron diagrams. The full set is necessary as an analysis of triple-Regge data finds that the triple-Pomeron coupling, $g_{3P}$, is three times larger than that coming from the original analyses, which did not include absorption. A novel feature of the model is that the Pomeron is described in terms of 3 interconnected components with different $k_t$, which allows the soft to hard Pomeron transition to be studied. The soft (small $k_t$) component is heavily screened, so that dependent observables, such as the total cross section and the parton multiplicity at low $k_t$, show relatively little growth with $\sqrt{s}$; $\sigma_{\rm tot}$ is prediced to be about 90 mb at the LHC energy $\sqrt{s}$=14 TeV. On the other hand, the large $k_t$ (or QCD) component suffers relatively little screening and the dependent observables have larger growth with $\sqrt{s}$.

It was emphasized that central exclusive production of a heavy system $A$, that is, the process $pp \to p+A+p$, offers an excellent opportunity to study the Higgs sector at the LHC via its $b\bar{b}$ decay channel, which, otherwise, is very challenging to observe. The crucial observation is that there exists a selection rule which greatly suppresses the $pp \to p+b\bar{b}+p$ QCD background. To identify the exclusive Higgs signal it is necessary to instal detectors for the outgoing very forward protons at a great distance (about 400 m) from the $pp$ interaction point. These will enable the Higgs mass to be measured via the missing mass method to an accuracy $O$(1) GeV. However, we must pay a price for the exclusivity of the Higgs signal. Soft rescattering can populate the rapidity gaps. We described how to calculate the survival probability of the gaps to both eikonal and enhanced rescattering. The latter, which involves rescattering on intermediate partons, breaks the soft-hard factorization of the amplitude for the process. Care must be taken to evaluate its effect. A vital step is the model of soft interactions with Pomeron exchange with different $k_t$ values, which enabled the $b \leftrightarrow k_t$ correlation to be taken into account. It was found that the suppression caused by enhanced rescattering is numerically not large for the exclusive production of a heavy mass system (like a Higgs boson) at the LHC. 

We noted two results of NLO analyses of HERA data. First, `flat' small-$x$ gluons are obtained at the relevant scales in the global fits of deep inelastic and related data. Second, small values of the saturation scale $Q_s(b)$ are obtained for impact parameters larger than about 0.6 fm. These observations imply that the gluon density, which gives rise to the enhanced absorptive correction, is not large, and that the black disc regime is not reached at the LHC\footnote{Note that the black disc regime of \cite{strik} was obtained using leading order (LO) gluons which grow steeply with $1/x$. This growth is simply an artefact of the absence of the $1/z$ singularity in the LO quark splitting function. When NLO contributions are included the deep inelastic data are described by flat (or even decreasing with $1/x$) gluons at the relevant low scales, see the discussion in \cite{RMK2}.}. The cross section predicted for a 120 GeV SM Higgs boson, before acceptance and $b\bar{b}$ identification cuts, is $2-3$ fb at the LHC energy $\sqrt{s}=14$ TeV; with a signal-to-background ratio of $O(1)$. The cross section and signal-to-background ratio may be up to an order-of-magnitude higher in some SUSY Higgs scenarios.

We detailed exclusive processes that have already been observed at the Tevatron with rates in agreement with predictions of the same model as that used for the Higgs estimates. This gives powerful support for the addition of forward proton taggers to enhance the discovery and physics potential of the ATLAS and CMS detectors at the LHC. Finally, we described a range of processes that may be observed in the early runs of the LHC which can provide valuable checks of all aspects of the theoretical formalism for exclusive processes.

\section*{Acknowledgements}

We thank Aliosha Kaidalov for useful discussions.
MGR thanks the IPPP at the University of Durham for hospitality.
The work was supported by  grant RFBR
07-02-00023, by the Russian State grant RSGSS-3628.2008.2.

\end{document}